\documentclass[11pt, oneside]{article}   	% use "amsart" instead of "article" for AMSLaTeX format
\usepackage{geometry}                		% See geometry.pdf to learn the layout options. There are lots.
\usepackage[parfill]{parskip}    		% Activate to begin paragraphs with an empty line rather than an indent
\usepackage{graphicx}				% Use pdf, png, jpg, or eps§ with pdflatex; use eps in DVI mode
\usepackage{amssymb}
\usepackage{amsmath}
\usepackage{dsfont}
\usepackage{enumitem}
\usepackage{authblk}
\usepackage{hyperref}

\newcommand{\er}{Erd\H{o}s-R\'enyi }

\newcommand{\avx}[1]{ {\langle{#1}\rangle} }
\newcommand{\av}[1]{ {\mathbb{E}\left[\,{#1}\,\right]} }
\newcommand{\avs}[1]{ \overline{{\mathbb{E}}}\left[\,{#1}\,\right]}
\newcommand{\var}[1]{ {\mathbb{V}\left[{#1}\right]} }
\newcommand{\vas}[1]{ \overline{{\mathbb{V}}}\left[{#1}\right] }
\newcommand{\avcd}[1]{\overline{#1}} % average over diagonal components
\newcommand{\avcm}[1]{\overline{\overline{#1}}} % average over matrix components
\newcommand{\tr}{ \textrm{tr} }
\newcommand{\id}{\mathbb{I}}
\newcommand{\one}{\mathbb{J}}

\newcommand{\gf}[1]{{\mathcal F}^{(#1)}}
\newcommand{\alp}{\alpha}
\newcommand{\mal}[1][ij]{\alpha_{#1}}
\newcommand{\aal}{{\bar \alpha}}
\newcommand{\sal}{\sigma_\alpha}
\newcommand{\amu}{\bar \mu}
\newcommand{\smu}{\sigma_\mu}
\newcommand{\ala}{\bar \Lambda}

\newcommand{\asi}{\bar \Sigma}
\newcommand{\ssi}{\sigma_\Sigma}

\newcommand{\veci}{\mathbf{i}}
\newcommand{\vecj}{\mathbf{j}}
\newcommand{\veca}{\mathbf{a}}
\newcommand{\vecz}{\mathbf{0}}
\newcommand{\veck}{\mathbf{k}}

\newlist{enh}{enumerate}{1}
\setlist[enh]{label=\bf{H\arabic*:}}

\title{Linear processes in high-dimension:\\ phase space and critical properties}
\author[1]{Iacopo Mastromatteo}
\author[1]{Emmanuel Bacry}
\author[1,2]{Jean-Fran\c{c}ois Muzy}
\affil[1]{\small Centre de Math\'ematiques Appliqu\'ees, CNRS, \'Ecole Polytechnique, \authorcr UMR 7641, 91128 Palaiseau, France}
\affil[2]{\small Laboratoire Sciences Pour l'Environnement, CNRS, Universit\'e de Corse, \authorcr UMR 6134, 20250 Cort\'e, France}
\date{}  % Activate to display a given date or no date

\begin{document}

\maketitle

\abstract{In this work we investigate the generic properties of a stochastic linear model in the regime of high-dimensionality. We consider in particular the Vector AutoRegressive model (VAR) and the multivariate Hawkes process. We analyze both deterministic and random versions of these models, showing the existence of a stable and an unstable phase. We find that along the transition region separating the two regimes, the correlations of the process decay slowly, and we characterize the conditions under which these slow correlations are expected to become power-laws. We check our findings with numerical simulations showing remarkable agreement with our predictions. We finally argue that real systems with a strong degree of self-interaction are naturally characterized by this type of slow relaxation of the correlations.}

\section{Introduction}
Stochastic linear models are instruments of paramount importance for describing physical, social or economic systems: despite being simple enough to be analytically tractable, they allow to accurately describe a large number of qualititatively different systems.
In this work, our focus will be the study of linear models in the regime of high dimensionality, and the analysis of the effects induced by the collective regime of interactions on the overall stability and on the first and second-order properties of the system.
Such analysis is related to multiple fields of activity previously considered in the literature:
\begin{itemize}
\item The properties of large ecosystems close to the equilibrium have been studied with similar methods, finding that universality controls the stability of large food webs~\cite{May:1972aa}. Our analysis extends in particular the results obtained in~\cite{Jirsa:2004aa}, which consider the \emph{dynamic} version of May's seminal model~\cite{May:1972aa}, to the case of a symmetric interaction kernel (see Sec.~\ref{sec:goe}).
\item Linear models for self-exciting marked point-processes are customarily used in order to model seismic activity \cite{Ogata:1988aa} (the mark being used to account for the intensity of the activity). This work generalizes this approach by suggesting that it is possible to see the components $i \in \{1,\dots,N\}$ as a set of \emph{interacting} marks. From this perspective, our results suggest that the critical relaxation measured in these systems might be endogenously generated by the interaction among a large number of marks.
\item Multivariate linear models are standard tools in order to account for the cross-correlation among economic variables~\cite{Sims:1980aa}, and in particular linear self-exciting point process can be used in order to assess the risk of contagion in a financial network~\cite{Aitsahalia:2010aa}.
\item Linearly interacting  point-process are used in order to model neural networks~\cite{Reynaud:2013aa}  made by a large number of components (although non-linear generalizations are customarily used in order to model inhibition).
\end{itemize}

\vskip .3cm

In all these cases, one is interested in characterizing the statistical properties of a system composed of a large number $N\gg 1$ of linearly interacting entities $X(t) = \{ X_i(t) \}_{i=1}^N$, $t$ being  interpreted as the time and $i$ as an index labeling the different entities.
A major concern is the limiting behavior of the system in the regime of high-dimensionality, as the limit $N\to\infty$ is expected to be fixed by universality.
We shall effectively address this issue through three angles, namely {\em Stability}, {\em Endogeneity} and {\em Relaxation}. Let us describe more clearly these perspectives:

{\bf Stability.} An exogenous noise can drive the system towards the instability point characterized by the divergence of one or more components of $X(t)$. Indeed, the interaction network generates feedback loops which may enhance the susceptivity of the system to external perturbations. Hence, we want to determine the critical interaction level beyond which such instability effects arise, dominating the behavior of the system. %We will refer to this angle as the one of {\bf Stability}.

{\bf Endogeneity.} We will be interested in characterizing what proportion of the \emph{average intensity}\footnote{
With a slight abuse of language, we will refer to $\Lambda$ as to the vector of average intensities for both processes that we will consider (VAR and Hawkes), see also the footnote of page~\pageref{fnt:abuse_not}.
}
  \begin{equation}
	\label{eq:2}
\Lambda_i=\frac{\langle dX_i(t)\rangle}{dt}
  \end{equation}
is related to exogenous factors and what proportion derives from endogenous cross-contamination effects among the components.

{\bf Relaxation.} As we will consider the components of the system to be individually characterized by short memory and fast relaxation, it will be interesting to study whether the lagged cross-correlations
\begin{equation}
  \label{eq:xcorr_intro}
  c_{ij}(\tau) = \frac{\avx{dX_i(t)dX_j(t+\tau)} -\avx{dX_i(t)}\avx{dX_j(t+\tau)}}{dt^2}
\end{equation}
display or not slow relaxation in the limit of large $\tau$ as a consequence of the collective regime of interactions.
\footnote{
While in principle one could quantify the amount of endogenous interaction by focusing on the cross-correlations, we choose to adopt the average intensity as a proxy. The results of Secs.~\ref{sec:det},~\ref{sec:goe} and~\ref{sec:rrg} show that both quantities appropriately signal the overall degree of self-interaction in the system.
}

Strongly endogenous effects are commonly found in applications along with slow relaxation of the auto-correlation (between the same components of $X(t)$) or the cross-correlation (between different components of $X(t)$) functions. This is, for instance, the case of high-frequency dynamics in financial markets.
A growing number of empirical works, describing by means of linear models this high-frequency dynamics, are in fact revealing that financial correlations are characterized by long-range memory, signaled by power law relaxations with small exponents \cite{Bacry:2012aa,Filimonov:2012aa,Hardiman:2013aa,Bacry:2013ac}.
This behavior can in principle be justified on the basis of the self-reflexive character of financial markets, by arguing that, as a consequence of the strongly endogenous regime in which markets operate, slow relaxation can be induced in the system. From this perspective, the results of Refs.~\cite{Bremaud:2001aa,Jaisson:2013aa} can be used in order to investigate analytically the onset of such a critical regime under the implicit assumptions that each of the entities composing the system is \emph{individually} poised at criticality.
Our aim is to discuss the complementary scenario in which long-range correlations in the system are induced \emph{as an effect of its interaction network}, by assuming that no component in the system singularly exhibits critical behavior.
More precisely, we shall suppose that individual units are fastly relaxing and try to
characterize the influence of the network structure on the long-time behavior ($\tau \rightarrow +\infty$) of the correlations $c_{ij}(\tau)$ in the limit $N \rightarrow +\infty$. The limit when the system can collectively encode \emph{long-memory}, signaled by the divergence of the quantity $\hat c_{ij}(0) = \int_0^\infty d\tau \,c_{ij}(\tau)$ will be referred to as the issue of {\bf Criticality} of the system.

Let us notice that, as a consequence of the assumption that individual units are fastly relaxing, the regime in which the $\tau\to\infty$ limit is taken before the $N\to\infty$ limit becomes trivial. On the other hand, taking the $N\to\infty$ limit before  any other limit
will allow us to access a very rich phenomenological behavior.

\vskip .2cm
The organization of the paper is the following. We introduce our main framework
for linear systems in continuous time
in Section~\ref{sec:lin_mod}. VAR models are first introduced, followed by Hawkes processes. Then, we restrict our study to the particular case of factorizable linear systems.
At the end of this Section, we present the methodology for analyzing these systems that will be used all along the paper on each model.%: Stability, Endogeneity, Relaxation and Criticality in the limit $N \rightarrow +\infty$.*** could we remove this part after the colon?***
Section~\ref{sec:det} studies the case a deterministic regular network of interactions whereas Sections~\ref{sec:goe} and~\ref{sec:rrg} study the case of random models within the framework of two tractable random matrix ensembles: the Gaussian Orthogonal Ensemble and the Regular Random Graph Ensemble. We draw our conclusions in Section~\ref{sec:conclusions}, while the more technical parts of the discussion are relegated to the Appendices.

\section{Linear systems in continuous time}
\label{sec:lin_mod}
In this preliminary section we define the classes of linear models on which we focus our analysis.  Beside fixing the notations and the conventions that we will follow throughout the paper, we aim to highlight the main similarities among the two linear models that will be presented. In particular, we want to show that the relations defining their behavior are formally identical and are essentially due
to their linear nature. More general linear models can be expected to share the same behavior, as long as relations such as~(\ref{eq:psi}),~(\ref{eq:av}) and~(\ref{eq:xcorr}) hold.

\subsection{Vector Autoregressive model}
The first type of model that we are going to consider is an $N$-dimensional Vector AutoRegressive Model (VAR), a widespread model introduced in~\cite{Sims:1980aa} and customarily used in econometrics in order to describe the dynamic relastionship among the different components of an economic time-series (for a more exhaustive account of the vast literature concerning VAR processes, we address the reader to the surveys~\cite{Hamilton:1994aa,Watson:1994aa}). The model is defined by a set of $N$ processes $X(t)=\{X_i(t)\}_{i=1}^N$ evolving in discrete time, driven by Gaussian noises $\eta(t) = \{ \eta_i(t) \}_{i=1}^N$,  and interacting through a linear matrix kernel $\Phi(\tau)$. In vector notation, the process is defined by the relation
\begin{equation}
\label{eq:ar}
X(t) =  \sum_{t^\prime=-\infty}^{t-1} \Phi(t-t^\prime) X(t^\prime) + \eta(t) \; ,
\end{equation}
where we assume the interaction kernel $\Phi(\tau)$ to admit the discrete Fourier transform
\begin{equation}
\label{eq:disc_fourier}
\hat \Phi(\omega) = \sum_{\tau=-\infty}^{+\infty} e^{-i\omega \tau} \Phi(\tau) \, ,
\end{equation}
with components $\hat \Phi(\omega) = \{ \hat \Phi_{ij}(\omega) \}_{ij=1}^N$.
We restrict ourselves to systems for which the interaction kernel $\Phi(\tau) $
is causal, i.e.,
\begin{equation}
\label{causality}
\Phi(\tau) = 0,~~~~ \forall \tau < 0.
\end{equation}

\vskip .3cm
\noindent
{\bf Stability of VAR models.}
Most of the time we shall consider that $\Phi$ satisfies the following so-called stability assumption :
\begin{itemize}
\item {\bf (H1) Stability Assumption.} The spectral radius of $\hat \Phi_0 $ is smaller than 1, which we can indicate by $|| \hat \Phi_0 || < 1$. Equivalently, all the eigenvalues of $\hat \Phi_0 $ have modulus smaller than one.
\end{itemize}
\vskip .2cm
Indeed, under this last assumption, it is possible to prove (see e.g.~\cite{Hamilton:1994aa}) that the infinite sum
\begin{equation}
\label{psi}
\Psi(\tau) = 1 + \Phi(\tau) + \Phi(\tau) *  \Phi(\tau) + \Phi(\tau) *  \Phi(\tau) *  \Phi(\tau) + \dots
\end{equation}
is well-defined, and its Fourier transform can be written as the matrix
\begin{equation}
\label{eq:psi}
\hat \Psi(\omega) = (\id - \hat \Phi(\omega))^{-1} \; ,
\end{equation}
where $\id$ denotes the $N$-dimensional identity matrix. The process $X(t)$ can then be written as the convolution
\begin{equation}
  \label{eq:conv_ar}
  X(t) = \Psi(t) * \eta(t) \; ,
\end{equation}
where, by convention, the $*$ operator refers to the regular matrix product where all multiplications have been replaced by discrete convolutions. The process $X(t)$ is proved to admit a stationary state, whose associated probability measure expectation will be denoted by the symbol $\avx{\dots}$. We will be interested in computing first and second order properties of the process under such measure whenever stability assumption hold.\\

\vskip .3cm
\noindent
{\bf Continuous time VAR models.}
In order to emphasize the analogy with a Hawkes process, in the next parts of the discussion we will be considering the continuous-time version of this model, also employed in~\cite{Potters:2005aa,Mastromatteo:2011aa} in order to model high-frequency financial data. In this case, Eq.~(\ref{eq:conv_ar}) can be generalized to the continuous case by replacing the discrete convolution with a continuous one. In particular, we can define the continuous-time VAR by substituting the Gaussian noises $\eta(t)$ with a Wiener process with increments $d\eta(t)$. The continuous time process satisfies, by construction, the same properties of its discrete-time counterpart once we define the continuous-time Fourier transform as
\begin{equation}
\label{eq:fourier}
\hat \Phi(\omega) = \int_{-\infty}^{+\infty} d\tau\, e^{-i\omega \tau} \Phi(\tau) \, .
\end{equation}
In particular, the analytical formulae for the average $\Lambda$ and the cross-correlation matrix $c(\tau)$ are given by the equations below. % average ``intensity'' $\Lambda$\footnote{Here, we use the term intensity as an analogy with point processes. This will be discussed at the end of Section \ref{sec:hawkesvar}.} and cross-correlation matrix are given by the equations below.

\vskip .3cm
\noindent
{\bf Endogeneity of VAR models.}
The expectation $\Lambda dt = \avx{dX(t)} $ can be expressed as
\begin{equation}
\label{eq:av}
\Lambda = \hat \Psi(0) \mu \; ,
\end{equation}
where we write $\avx{d\eta(t) } = \mu \, dt $.
The relation among endogenous and exogenous effects is determined by the matrix $\Psi(0)$, whose spectral norm determines the maximum output intensity $\Lambda$ as a function of the norm of the driving vector $\mu$.

\vskip .3cm
\noindent
{\bf Correlation of VAR models.}
The cross-correlation matrix $$c(t-t^\prime) dt dt^\prime = \avx{ dX(t) dX^T(t^\prime) } -  \avx{ dX(t) } \avx{ dX^T(t^\prime) }$$ can be obtained through the Fourier transform $\hat c(\omega)$, which is given by
\begin{equation}
\label{eq:xcorr}
\hat c(\omega) = \hat \Psi^\star(\omega) \Sigma \hat \Psi^T(\omega) \; ,
\end{equation}
where the symbol $^T$ denotes matrix transposition and $^\star$ indicates complex conjugation.
We have assumed the covariance term $\Sigma$ to be defined by the relation
 \begin{equation}
 \label{sigmavar}
 dt dt^\prime \,\Sigma \delta(t-t^\prime)
 =\avx{ d\eta(t) d\eta^T(t^\prime) } - \avx{ d\eta(t)} \avx{ d\eta^T(t^\prime) },
 \end{equation}
 where
 $\delta(\tau)$ indicates the Dirac delta function. Moreover, we assumed  the matrix $\Sigma$ to be diagonal.
 %Here, and in the following,  , while we will employ the notation $\delta_{ij}$ for the Kronecker symbol, equal to one if $i=j$  and zero otherwise.

\subsection{Hawkes process}
The second class of models that we consider are Hawkes processes, a class of interacting point processes customarily used to describe self and cross-excitation phenomena \cite{Hawkes:1971lc,Hawkes:1971nq}.
For a long time,
Hawkes models have been extensively used to
describe the occurrence of earthquakes in some given region \cite{ogata99,helsor02}.
They are getting more and more popular in many other applications
in which tracking how information diffuses through different ``agents'' is the main concern, e.g.,
neurobiology (neurons activity) \cite{Rey2},  sociology (spread of terrorist activity) \cite{emhawkes,mo11} or processes on the internet (viral diffusion across social networks) \cite{cranesor08,yangzha}.
Their application to finance can be traced back to Refs~\cite{Chavez-Demoulin:2005aa,Hewlett:2006aa}, and has been followed by a still-ongoing spree of activity~\cite{Bowsher:2007aa,Bauwens:2009aa,Giesecke:2011aa,Toke:2011aa,Embrechts:2011aa,Bacry:2012aa,Filimonov:2012aa,Bacry:2013aa,Bacry:2013ab,Bacry:2013ac,Hardiman:2013aa}.

An $N$-dimensional Hawkes process is defined by a set of $N$ counting processes evolving in continuous time $X(t) = \{ X_i(t) \}_{i=1}^N$\footnote{\label{fnt:abuse_not}With abuse of notation, we are denoting quantities describing the VAR model with the same symbols adopted for the Hawkes process. We choose to do so in order to show more transparently the close relation among the two models, which satisfy extremely similar relations. We will specify explicitly which of the two frameworks we are considering whenever this notation results ambiguous.}. The probability for an event to be triggered is expressed by a stochastic intensity function $\Lambda(t)=\{\Lambda_i(t) \}_{i=1}^N$ which evolves according to the dynamics:
\label{eq:ar_cont}
\begin{equation}
\Lambda(t) = \mu + \int_{-\infty}^{t} \Phi(t-t^\prime) dX(t^\prime) \; ,
\end{equation}
where the components of $\mu$ are commonly referred as \emph{exogenous intensities} (or {\em baseline intensities}), and $\Phi(t)$ is a positive semidefinite, causal (in the sense of \eqref{causality}), locally $L^1$ matrix kernel. Notice that, unlike in the VAR case,
in order for the event probabilities $\Lambda(t)dt$ to be well-defined,
we need to assume that
\begin{equation}
\nonumber
\Phi(\tau) {\mbox {~~~satisfies component-wise positivity}}.
\end{equation}
{\bf Stability of Hawkes processes.}
If a Hawkes process satisfies the stability assumption {\bf (H1)} specified above for the VAR model, then one can show that $X(t)$ is stationary and stable \cite{Hawkes:1971lc,Hawkes:1971nq}. Moreover, as in the VAR model, this condition implies that
the infinite sum $\Psi(\tau)$ (Eq. \eqref{psi}) is well-defined, and its Fourier transform $\hat \Psi(\omega)$ is given by Eq.~(\ref{eq:psi}).

\vskip .3cm
\noindent
{\bf Endogeneity of Hawkes processes.} As for the VAR model,
the mean intensity $\avx{ \Lambda(t) } dt = \avx{ dX(t) } = \Lambda\, dt$ is expressed by Eq.~(\ref{eq:av}). Again, the ratio of endogeneity versus exogeneity is is determined by the matrix $\Psi(0)$, which sets the relation among endogenous intensity $\Lambda$ and exogenous intensity $\mu$.

\vskip .3cm
\noindent
{\bf Correlation of Hawkes processes.} Again, as for the VAR model,
the cross-correlation matrix $$c(t-t^\prime) dt dt^\prime=\avx{ dX(t) dX^T(t^\prime) } -  \avx{ dX(t) } \avx{ dX^T(t^\prime) }$$ can be expressed by Eq.~(\ref{eq:xcorr}) in which $\Sigma$
is the diagonal matrix defined by (see~\cite{Bacry:2012aa})
\begin{equation}
 \label{sigmahawkes}
\Sigma_{ij}=\Lambda_i \delta_{ij},
\end{equation}
where $\delta_{ij}$ stands for the Kronecker symbol, equal to one if $i=j$  and zero otherwise.

\subsection{Hawkes processes versus VAR processes.}
\label{sec:hawkesvar}
The previous results indicate that a Hawkes process is very reminiscent of a VAR process formulated in continuous time with non-negative $\Phi(\tau)$, although some important differences need to be emphasized.

 First, while the $\mu$ in the autoregressive case identify the average increments of a multivariate Wiener process, in the Hawkes case they emerge as the exogenous component of the average intensity.

 Secondly, in the autoregressive case the covariance of the noise $\Sigma$ controlling the cross-correlations is independent of $\Phi(\tau)$ and $\mu$, while in the Hawkes case it is \emph{endogenously generated} and is thus fixed to the values $\Sigma_{ii}=\Lambda_i$.

 Finally, the cross-correlation $c(\tau)$ of a Hawkes process is always singular: as we are considering counting processes with jumps of size one, the relation $dX_i(t) =(dX_i(t))^2$ holds. Then, the cross correlations can be decomposed as $c_{ij}(\tau) = \Lambda_i \delta_{ij}\delta(\tau) + c^{(reg)}_{ij}(\tau)$ where $c^{(reg)}_{ij}(\tau)$ is regular around zero. For VAR processes, such a singular contribution to the correlations emerges just if $\avx{ d\eta(t) d\eta^T(t^\prime) } - \avx{ d\eta(t)} \avx{ d\eta^T(t^\prime) }$ contains a singular component centered at zero.

Let us point out that, with a slight abuse of language, we will refer to  $\Lambda$ and  $\mu$ respectively as mean intensities and exogenous intensities even in the case of a VAR model.

\subsection{Factorizable linear systems}
\subsubsection{An isotropical factorizable interaction kernel}
\label{iso}
As we are interested in studying how heterogeneity in the dynamic behavior of the system may emerge as an effect of the interaction network, we take models for which the interaction kernel $\Phi(\tau)$ can be factorized as
\begin{equation}
\label{eq:factasumpt}
\Phi(\tau) = \alp \phi(\tau) \; ,
\end{equation}
where $\alpha$ is a matrix and $\phi(\tau)$ is a scalar function of $\tau$ satisfying (without loss of generality) $$\hat \phi(0) = \int_0^{\infty} d\tau \phi(\tau)= 1.$$ We additionally suppose that $$ \int_0^{\infty} d\tau \, \tau \phi(\tau) < \infty,$$ as we want to focus on the problem of whether long-range memory in correlations can be \emph{endogenously} induced by a system whose interactions have short-range memory. Hence, we are supposing the different components of the system to react homogeneously in time (i.e., with the same speed) to innovations, while we confine the heterogeneity of the system to the interaction strengths, which are specified by the matrix $\alp$. Moreover, the $\alp$ matrix is naturally interpreted as a weighted adjacency matrix specifying the overall strength of the interaction between the different components of the system\footnote{Let us point a factorization assumption like~(\ref{eq:factasumpt}) is very common in space-time Hawkes models for earthquakes~\cite{ogata99}.}.  We will consider exclusively  cases for which the matrix $\alp$ can be diagonalized, so that we will always be able to write
\begin{equation}
\label{eq:decomp}
\alp =  U  \lambda U^{-1} \; ,
\end{equation}
where $\lambda$ is a diagonal matrix of eigenvalues with elements equal to $ \{ \lambda_a\}_{a=1}^N$ and $U$ is a suitable change of basis matrix, with entries denoted by $\{U_{ia}\}_{ia=1}^N$. Note that we are adopting the convention of using the indices $i,j,\dots$ to denote the components of $X(t)$ and $a,b,\dots$ for the eigenvalues of the $\alp$ matrix.
For simplicity, We will further require some supplementary conditions to hold for the system, namely
\vskip .1cm
\begin{itemize}[resume]
\item {\bf (H2) Unitarity Assumption.} The matrix $U$ is assumed to be unitary (i.e. $U^\dagger=U^{-1}$, where $^\dagger$ denotes Hermitian conjugation).
\item {\bf (H3) Homogeneity Assumption.} All the components of $\Sigma$ (defined by \eqref{sigmavar} in the case of the VAR model and by \eqref{sigmahawkes} in the case of the Hawkes model) are assumed to be equal to $\asi$, and the components of the the mean intensity vector are equal to $\ala$.\footnote{While in the translationally invariant case discussed in Sec.~\ref{sec:det} the system will fulfill by construction the homogeneity assumption defined above, in the random cases presented in Secs.~\ref{sec:goe} and~\ref{sec:rrg} this assumption will hold on average, allowing the system to enjoy the same properties.}
\end{itemize}

\noindent
\vskip 0.3cm
The diagonalization assumption \eqref{eq:decomp} allows to write the components of the matrix $\hat \Psi(\omega)$ as
\begin{eqnarray}
  \label{eq:decomp_psi}
  \hat \Psi(\omega) = \hat \phi(\omega)^{-1} U \left( \hat \phi(\omega)^{-1} - \lambda\right)^{-1} U^{-1} \; .
\end{eqnarray}
We will often employ this decomposition of $\hat\Psi(\omega)$ in order to link the distribution of eigenvalues of the matrix $\alp$ with the value of the observables $\Lambda$ and $c(\omega)$.

The unitarity {\bf (H2)} and homogeneity {\bf (H3)} assumptions above
allow us to disregard the effect on the system of the heterogeneity in the angular components of the interaction matrix, permitting us to focus on the collective effect induced by the isotropical part of the interaction.
Even though the global effect of the inhomogeneity of the system is an interesting problem on his own, the simpler case that we consider is a necessary first step in order to understand the collective effects of the large $N$ limit on this type of systems.

\vskip .3cm
In the homogeneous case, it will be useful to introduce the notation
\begin{eqnarray}
  \avcd{\Lambda} &=&\frac 1 N \sum_i \Lambda_i \\
  \avcd{c(\tau)} &=& \frac 1 N \sum_i c_{ii}(\tau)  \\
   \avcm{c(\tau)}&=& \frac 1 {N^2} \sum_{ij} c_{ij}(\tau) \;,
\end{eqnarray}
as all the information about the vector of average intensity and the lagged cross-correlation matrix is encoded in these averages over components.

Under the unitarity {\bf (H2)} and homogeneity {\bf (H3)}  assumptions, it is easy to show that Eq.~(\ref{eq:xcorr}) leads to
\begin{equation}
  \label{eq:acorr_factor}
  \avcd{\hat c(\omega)} = \frac 1 N \sum_i \hat c_{ii}(\omega) = \frac{\asi}{N} \sum_a \frac 1 {||\hat \phi(\omega)^{-1} -\lambda_a ||^2} \; ,
\end{equation}
so that the information about the eigenvectors encoded in $U$ is not required in order to understand the behavior of the average over components of the autocorrelation function.
{\bf The behavior of autocorrelations is completely specified through the function $\phi(t)$ and the spectrum of the matrix $\alp$}.

\subsubsection{Case of an exponential factorizable kernel}
Finally, we want to introduce an useful benchmark case for the kernel function $\phi(\tau)$, which will be employed in order to characterize the long-time behavior of the system. Specifically, we will often particularize our results to the special case in which the interaction kernel has an exponential shape of the form $$\phi(\tau) = \beta e^{-\beta \tau} \mathds{1}_{\mathbb{R}^+}(\tau),$$ or equivalently $$\hat \phi(\omega) = (1 + i \omega/\beta)^{-1},$$ where $\mathds{1}_{\mathbb{R}^+}(\tau)$ denotes the indicator function on $\mathbb{R}^+$. This particularly simple case will allow us to explore the main qualitative features of the results without losing analytical control on the solution.
In this framework, the locations of the poles appearing in the Fourier transforms of the correlation function~(\ref{eq:xcorr}) are trivially dictated by the eigenvalues of the matrix $\alp$. The poles can be written as
\begin{equation}
\label{eq:poles}
\omega_a = i \beta (1-\lambda_a) \quad \quad \omega_a^\star = -i \beta (1-\lambda_a^\star) \; ,
\end{equation}
where the $\{\lambda_a\}_{a=1}^N$ denote as usual the eigenvalues of $\alp$.

\vskip .3cm
\noindent
{\bf Relaxation.}
In the exponential case, inverting the Fourier transforms appearing in Eq.~(\ref{eq:xcorr}) becomes straightforward, if one supposes the eigenvalues to be non-degenerate. One obtains in fact
\begin{eqnarray}
\label{eq:xcorr_real}
c_{ij}(\tau) &=& \beta \theta(-\tau) \sum_{abk}  \frac{\lambda^\star_a (2-\lambda^\star_a) }{2-\lambda_b -\lambda^\star_a} U^\star_{ia} U^{\star \,-1}_{ak} \Sigma_k U^{-1}_{bk} U_{jb} e^{\beta [1-\lambda^\star_a]\tau} \\
&+& \beta \theta(\phantom{-}\tau) \sum_{abk} \frac{\lambda_b (2-\lambda_b) }{2-\lambda_b -\lambda^\star_a} \nonumber
 U^\star_{ia}  U^{\star \,-1}_{ak} \Sigma_k U^{-1}_{bk} U_{jb} e^{-\beta [1-\lambda_b]\tau} \; .
\end{eqnarray}
Under {\bf (H2)} and {\bf (H3)}, above equation leads to a simplified form of the average autocorrelation, which -- just as Eq.~(\ref{eq:acorr_factor}) above -- is independent of the eigenvectors of $\alp$:
\begin{eqnarray}
\label{eq:acorr_factor_real}
\avcd{c(\tau)} = \frac 1 N \sum_i c_{ii}(\tau)
&=&
\frac{ \beta \bar \Sigma}{N} \sum_a  \frac{\lambda_a (2-\lambda_a) }{(2-\lambda_a-\lambda_a^\star )} e^{-\beta [1-\lambda_a] |\tau|}  \; ,
\end{eqnarray}

Each eigenvalue of $\alp$ generates a decay mode indexed by $a$ whose associated speed depends upon the distance between the corresponding $\lambda_a$ and 1. In particular, the slowest mode $\lambda_{max} < 1$ controls the behavior of the correlations at large times, fixing the scaling of the correlation at large times to be proportional to $c_{ii}(\tau) \sim \exp (-\beta (1-\lambda_{max})|\tau|)$.  The \emph{critical} regime in which the support of the distribution of $\lambda_a$ touches the instability point $\lambda=1$ is of particular interest, and will be considered in specific sections (Secs.~\ref{sec:crit_dim_d},~\ref{sec:crit_dim_1},~\ref{sec:crit_goe} and~\ref{sec:crit_rrg}).

\vskip .3cm
\noindent
{\bf Genericity of exponential kernel.}
The main interest in considering the exponential kernel lies in the fact that it allows one to explore the long-time behavior of \emph{any} short-range kernel.
In fact, consider a generic kernel satisfying the \emph{short-range} assumption
\begin{equation}
  \label{eq:short_range_kern}
  \int_0^\infty d\tau \, \tau \phi(\tau) < \infty \; .
\end{equation}
This condition corresponds in Fourier space to the differentiability in zero of the function $\hat \phi(\omega)$, implying that it is possible to expand it as
\begin{equation}
  \label{eq:short_range_four}
  \hat \phi(\omega)= 1 + i\omega/\beta + o(\omega) \; ,
\end{equation}
with $\lim_{\omega\to 0} o(\omega)/\omega = 0$.
By back-transforming Eq.~(\ref{eq:xcorr}) in real space while keeping into account this expansion, one finds that the leading order term of the large time expansion of the cross-correlations is given by Eq.~(\ref{eq:xcorr_real}).
Intuitively, as the limit of large times $\tau\to\infty$ corresponds to the small $\omega$ regime in Fourier space, this indicates that at large times one can neglect the $o(\omega)$ term in the expansion of the kernel, sticking with the term $1 + i\omega/\beta$ which corresponds in real space to the exponential kernel
\begin{equation}
  \label{eq:exp_kern}
  \phi(\tau) = \beta e^{-\beta \tau}\mathds{1}_{\mathbb{R}^+}(\tau) \; .
\end{equation}

\subsection{Methodology for analyzing factorizable linear systems}
The next sections will analyze the behavior of factorizable linear systems (as defined in Section \ref{iso}), addressing systematically, as explained in the introduction, the issues of stability, endogeneity and  relaxation by focusing on a small set of relevant scalars encoding the collective information about the state of the system. In order to do so, we shall study the behavior of different observables.
More precisely :
\vskip .1cm
\begin{itemize}
\item {\bf Stability: The spectral norm of $\alpha$.} Due to Eq.~(\ref{eq:av}), the value of the average intensities $\Lambda_i$ for a fixed value of the input $\mu_i$ is determined by the largest eigenvalue of the matrix $\hat \Psi(0)$, which is precisely equal to $1/(1-||\alpha||)$. In particular, for $||\alpha||=1$ (i.e., the largest eigenvalue of $\alpha$ is equal to 1), endogenous effects spoil the validity of {\bf (H1)}, compromising the stationarity of the system and inducing a divergence of the averages.
\item {\bf Endogeneity: $\avcd{\Lambda}$.} We will be able to express $\avcd{\Lambda}$ as a linear function of $\avcd{\mu}$ thanks to the homogeneity hypothesis {\bf (H3)}, encoding the information about $\Psi(0)$ in the ratio $ \avcd{\Lambda}/\avcd{\mu}$. Hence, we will be able to interpret $\avcd{\Lambda}/\avcd{\mu}$ as a susceptibility of the system with respect to its driving input, while $\avcd{\Lambda}/\avcd{\mu}-1$ can be thought of as expressing the ratio between the endogenous intensity against the exogenous one.
\item {\bf Relaxation: $\avcd{ c(\tau)}$ and $\avcm{c(\tau)}$.} Slow relaxation will be characterized by the behavior at large times of the auto-correlations $\avcd{ c(\tau)}$ and of the cross-correlations $\avcm{c(\tau)}$, distinguishing the cases in which these functions decay exponentially from the one in which they develop broad tails. These quantities can be thought as quantifying the response of the system to noise, as it is proportional to the noise covariance matrix $\Sigma$.
\end{itemize}

All these quantities will be analyzed in the $N\to\infty$ regime of large dimensionality, in which non-trivial dynamic effects are expected to emerge. We shall focus in particular on the issue of {\bf Criticality}, which may induce long-range correlations (divergence of the $L^1$ norm of the self-correlation function).

Accordingly, the scaling in $N$ that we have introduced for the above observables is such that they all attain a finite limit for $N\to\infty$.  Nevertheless, specifying how such limit is approached requires a prescription encoding the dependence upon $N$ of the interaction kernels $\Phi(\tau)$, of the components $\mu$ and, if needed, of the parameter $\Sigma$.
In the next section we will use a deterministic prescription in order to scale the system with $N$, and we will explore the large $N$ regime of a regular lattice of finite dimension (i.e., $\alp$ is invariant under translation and the coefficients $\mu$ and $\Sigma$ are constants). Only in Secs.~\ref{sec:goe} and~\ref{sec:rrg} we will study the case in which the coefficients defining the model are free to fluctuate within two given statistical ensembles.

\section{The deterministic case}
\label{sec:det}
\subsection{A translationally-invariant network}
In this section we will discuss the case of a translationally-invariant network of interactions for the process $X(t)$, whose components are arranged on the vertices of a regular lattice. This prescription is used in order to explore the effect of the interactions in a completely homogeneous network, disregarding the  presence of irregularities in the system. Moreover, this framework  allows  to discuss the effect of the variation of the connectivity of the system, interpolating from the complete network (corresponding to the infinite dimensional limit, in which the interactions are broadly diluted through the system) to the low-dimensional case (in which strong fluctuations are induced as an effect of the topology), with the main advantage of controlling analytically the behavior of the system throughout the crossover.

\begin{figure}
  \centering
  \includegraphics[width=2in]{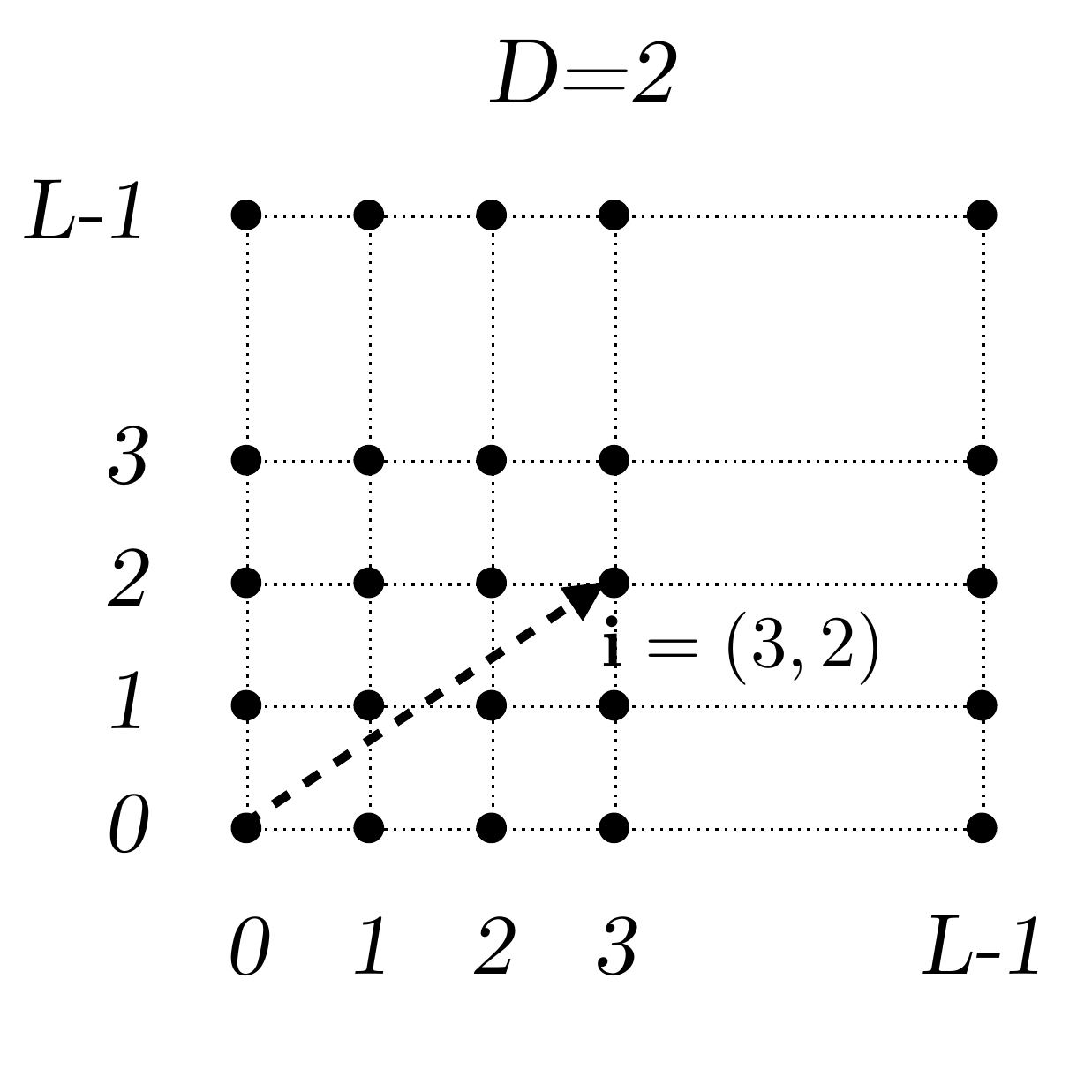}
  \caption{Sketch of a translationally-invariant network for $D=2$. In this case the components of $X(t)$ are arranged on the vertices of a square lattice of length $L$.}
  \label{fig:2d}
\end{figure}

In order to define the notion of spatial dimension, we preliminarly need to assume that a notion of geometry emerges naturally in the system, as its structure needs to be invariant under translations along a set of $D$ directions. In particular, we consider the case in which each of the components is located on the vertices of a hypercube of dimension $D>0$ (see Fig.~\ref{fig:2d}).
We will further assume the coordinates to have length $L$, so that we will be able to label them by using a vector index $\veci \in \{0,\dots , L-1\}^D$, and to identify the size of the system $N$ with the volume of the hypercube $N=L^D$.
\footnote{These models can be thought as discretized versions of continuous-space ones such as the one employed in~\cite{Ogata:1998aa} in order to model the occurrence of earthquakes. In our language the continuous-space limit is recovered in the limit $N\to\infty$. Moreover, the results shown in Sec.~\ref{sec:crit_dim_d} allow us to explore the regime in which long-range behavior is endogenously induced from the structure of the interactions, rather than enforced by construction in the parametric specification of the interaction model as it is assumed in~\cite{Ogata:1998aa}.
}
We will finally assume $\mu_{\veci}$ and $\Sigma_{\veci\veci}$ to be constants, i.e.,
$$\mu_{\veci} = \bar \mu ~~{\mbox {and}}~~ \Sigma_{\veci\veci} = \bar \Sigma,~~~\forall\,\veci \in [0,L-1]^D.$$
The periodicity condition will be enforced through the assumption
$$ \alp_{\veci \vecj}=\alp_{\veci - \vecj}~~~\forall \, \veci,\vecj \in [0,L-1]^D.$$

Let us point out that the diagonalization assumption \eqref{eq:decomp} as well as the Unitarity assumption {\bf (H2)} and the Homogeneity assumption {\bf (H3)} are all satisfied. Indeed, straightforward computations lead to
\begin{eqnarray}
  \label{eq:periodic_decomp}
  U_{\veci \veca} &=&  \frac{1}{\sqrt{N}} e^{\frac{2\pi i}{L} \veci \cdot \veca } \; ,\\
  \lambda_{\veca} &=&  \sum_{\veci} e^{\frac{2\pi i}{L} \veci \cdot \veca } \alpha_{\veci}\;,
\end{eqnarray}
where $\veca \in \{0,\dots, L-1\}^D$.

\subsection{Stability of translationally invariant systems}
The condition of stability of the system (i.e.\ assumption {\bf (H1)}) is equivalent to the condition
$$||\alpha|| = \max_\veca | \lambda_\veca | <1$$
which we shall assume to be fulfilled both in the VAR and in the Hawkes case.
In the latter case, or in the VAR model when all the entries of $\alpha$ are positive,
such condition simplifies to  $\lambda_{\vecz}< 1$.
Let us point out that, in this case, the stability of the system is controlled by the parameter $||\alpha||=\lambda_{\vecz}$, which increases the degree of self-interaction of the system. In particular, a strongly susceptible behavior is detected in the regime of $\lambda_{\vecz}$ close to 1, where the ratio $\avcd{\Lambda}/\avcd{\mu} -1 $ of endogenously-generated versus exogenously-generated intensity becomes extremely large (see~\eqref{eq:mean_lambda_lattice}).

\subsection{Endogeneity and Relaxation of translationally invariant systems}
One can easily prove that
\begin{eqnarray}
  \label{eq:mean_lambda_lattice}
  \ala &=& \frac {\amu} {1 - \lambda_{\vecz}} \\
  \avcd{\hat c(\omega)} &=& \frac {\asi} N ||\hat \phi(\omega)||^{-2} \sum_{\veca} \frac 1 {||\hat \phi(\omega)^{-1} - \lambda_{\veca}||^2 } \\
  \avcm{ \hat c(\omega)} &=& \frac {\asi} N ||\hat \phi(\omega)||^{-2}\frac 1 {||\hat \phi(\omega)^{-1} - \lambda_{\vecz}||^2 } \; .
\end{eqnarray}
Also notice that the expressions of $\ala$ and $\avcm{ \hat c(\omega)}$ depend just upon the single eigenvalue $\lambda_\vecz$. This is because the corresponding eigenvalue is $U_{\veci \vecz }=N^{-1/2}(1,\dots,1)$ so that by orthogonality one has $\sum_\veci U_{\veci \veca}=\sqrt{N} \delta_{\veca \vecz}$. An identical phenomenon will occur in Sec.~\ref{sec:rrg}, due to the absence of fluctuations in the sum over row of the $\alp$ matrix.\\

The autocorrelation function is more conveniently analyzed by rewriting it as
\begin{equation}
  \label{eq:ac_spectrum}
  \avcd{ \hat c(\omega)} = \asi\, ||\hat \phi(\omega)||^{-2} \int_{-\infty}^{+\infty} d\lambda\, \frac { \rho(\lambda)} {||\hat \phi(\omega)^{-1} - \lambda||^2}\; ,
\end{equation}
after introducing the \emph{spectral density}
\begin{equation}
  \label{eq:spectrum}
  \rho(\lambda) = \frac 1 N \sum_{\veca} \delta(\lambda - \lambda_{\veca}) \; .
\end{equation}
The advantage of introducing the spectral density is that $\rho(\lambda)$ is expected to have a well-behaved limit for $L\to\infty$, which we can exploit  in order to find the shape of correlations in the large size limit. In particular, if the maximum of the support of the spectrum is $\lambda_{max}<1$, the auto-correlations decay exponentially at large times, while for $\lambda_{max}=1$, the behavior of the spectrum close to $\lambda=1$ may lead to a non-exponential behavior of the correlation in the $L\to\infty$ limit.
These scenarios are illustrated by mean of some examples in the next Section.

\subsection{Criticality for a non-negative $\alp$}
\label{sec:crit_dim_d}
We first want to investigate the behavior of the model in the vicinity of $\lambda_{max} = 1$ in the case in which $\alpha_\veci$ is non-negative for all $\veci$.
Eq.~(\ref{eq:mean_lambda_lattice}) indicates that, for $\lambda_{max}\to 1$, the average intensity $\avcd{\Lambda}$ diverges, and the ratio of endogenous-to-exogenous events explodes. This type of divergence will be common to all the critical cases which we will be investigating, and can be reabsorbed into a suitable definition of $\avcd{\mu}$, similar to what is done in~\cite{Bremaud:2001aa}\footnote{Unlike in~\cite{Bremaud:2001aa}, where it is proven that the existence of a well-behaved $\lambda_{max}\to 1$ limit for the correlation in one dimension requires the kernel $\phi(\tau)$ to be long-ranged, we find that in the multidimensional setting even a short-ranged $\phi(\tau)$ may lead to a well-behaved limit for the correlations. Intuitively, for a fixed value of $\lambda_{max}$, a more densely wired system can redistribute potentially dangerous fluctuations among the components of the system, eventually avoiding the divergence of $c_{ii}(\tau)$.}.
Physically, this corresponds to the fact that a critical system operates in a regime where small input signals are translated into large outputs.

Interestingly, the average autocorrelation $\avcd{c(\tau)}$ can have a finite limit even though the averge intensity does not.
Calculating it requires calculating the behavior of the spectrum $\lambda_{\veca}$ close to the maximum eigenvalue, which is equal to $||\alpha||=\lambda_{\vecz}$.
Then we can set $\veck=2\pi \veca/L$, and calculate the limit
\begin{equation}
 \lambda(\veck)=\lim_{L\to\infty}\lambda_{\frac{\veck L}{2\pi}} \;,
\end{equation}
which is maximal for  $\veck=\vecz$, so that the system is stable for $\lambda_{max} = \lambda(\vecz) < 1$.

\subsubsection{Case of a parabolic spectrum}
\label{sec:parabolic}
 If the function $\lambda(\veck)$ is twice differentiable around zero, then its gradient vanishes and its Hessian is negative definite. In this case it is easy to estimate (see App.~\ref{app:fin_dim}) the $L\to\infty$ limit of the density $\rho(\lambda)$ close to the point $\lambda=\lambda(\vecz)$, which reads
\begin{equation}
  \label{eq:density_multidim}
  \rho(\lambda(\vecz)-\epsilon) \approx \det (-H[\lambda(\vecz)])^{-1/2} \frac{\epsilon^{D/2-1} }{\Gamma(D/2)}\left( \frac 1 {2\pi} \right)^{D/2} \;,
\end{equation}
where $H[\lambda(\vecz)]$ is the Hessian of $\lambda(\veck)$ calculated in $\veck=\vecz$.
Eq.~(\ref{eq:acorr_factor_real}) allows us to relate the exponent of $\epsilon$ in above expansion to the limiting behavior of the autocorrelations when $\bar \alpha$ approaches 1. In particular, the autocorrelations diverge approaching the instability point for $D=1,2$, while for $D>2$ they result
\begin{equation}
  \label{eq:acorr_multidim}
  \avcd{ c(\tau)} \sim \tau^{1-D/2} \;.
\end{equation}
This implies that
\begin{itemize}
\item For $D=3,4$ the process can develop long range memory when $\lambda(\vecz)$ tends to the instability point $\lambda(\vecz)=1$.
\item For $D\to\infty$ one finds instead that the system loses its power-law behavior, as the tails of the correlations become increasingly steeper.
\end{itemize}
Notice that the above result does not depend from the specific form of the interaction that we have chosen, but simply emerges from the non-negativity of $\alpha_\veci$ and the differentiability of $\lambda(\veck)$ around zero.

\vskip .3cm
\noindent
{\bf An example: next-neighbor interaction.}
As an example, one can study the behavior of a $D$-dimensional system with next-neighbor interactions, defined through $\alpha_{\veci}=\aal (2D)^{-1} \sum_{d=1}^D \left( \delta_{\veci - \mathbf{e}_d}+\delta_{\veci + \mathbf{e}_d}\right)$, with $\mathbf{e}_d$ denoting the $d$-th element of the canonical basis of $\mathbb{R}^D$. The spectrum of such a system is given by
  \begin{equation}
	\label{eq:spectrum_multidim}
\lambda(\veck) = \frac{\aal}{D} \sum_{d=1}^D \cos\left( k_d \right) \; .
  \end{equation}
For such functional form of $\lambda(\veck)$ one can explicitly calculate the limiting value of the spectral density $\rho(\lambda)$ for $N\to\infty$ (App.~\ref{app:fin_dim}), which results
\begin{equation}
  \label{eq:spec}
  \rho(\lambda) = \frac D {\aal} \int_{-\infty}^{+\infty} \frac{dz}{2\pi} \, e^{iz\lambda D/\aal} J_0^D(z)
\end{equation}
where $J_n(z)$ is the Bessel function of the first kind of order $n$ calculated in $z$. One can verify by expanding $\rho(\lambda)$ close to the point $\lambda=\aal = 1$ that its limiting behavior is described by Eq.~(\ref{eq:density_multidim}). Indeed, the procedure that we have given simply requires the calculation of $\lambda(\veck)$ around $\veck = \vecz$, and hence can be used in order to tackle a larger class of problems.

\begin{figure}
  \centering
  \includegraphics{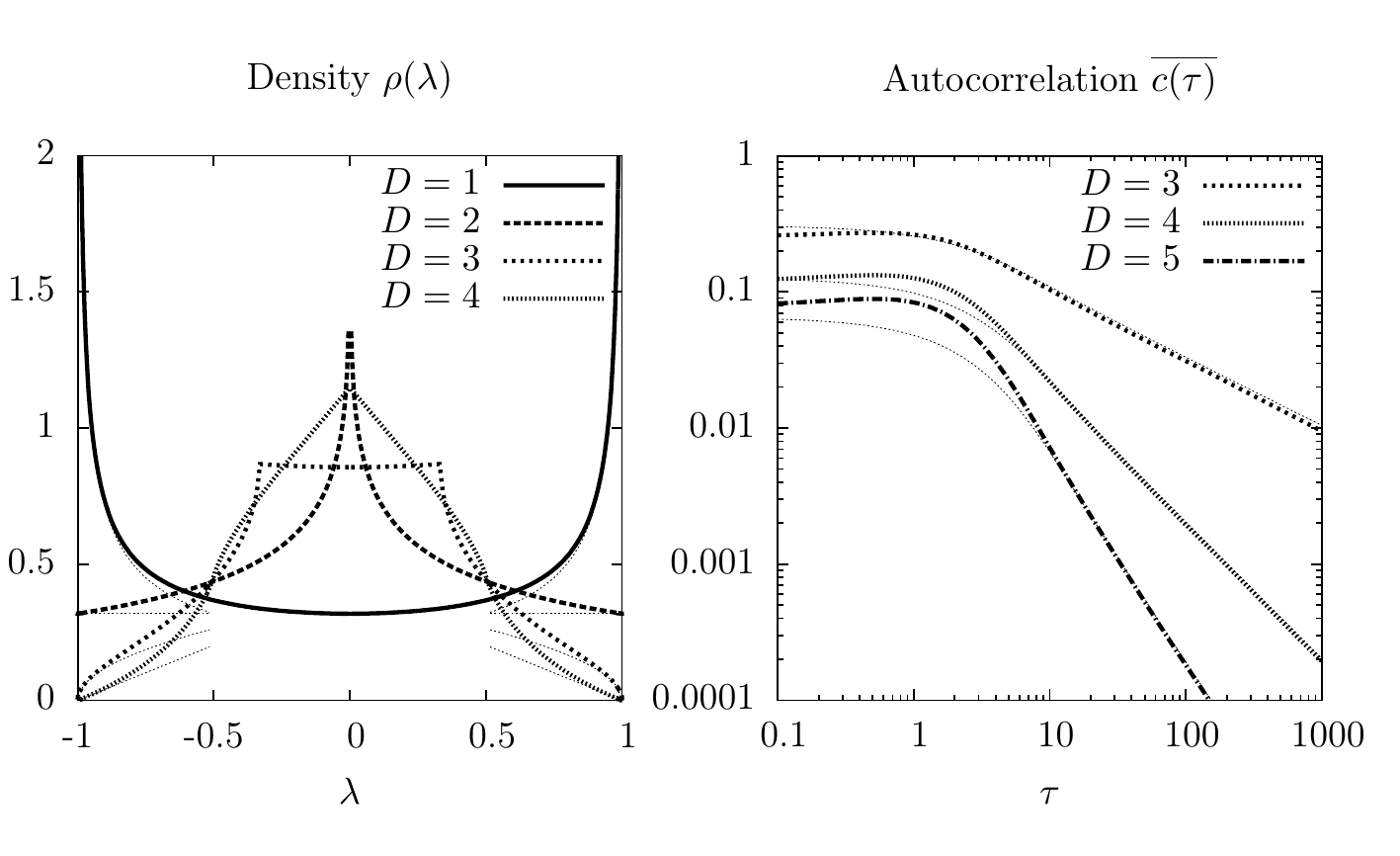}
  \caption{\emph{(Left panel)} Spectrum $\rho(\lambda)$ of a translationally invariant system with next-neighbor interactions for an interaction strength $\aal=1$ and different values of the dimension $D$ (see Example of Section \ref{sec:parabolic}). While for $D=1$ the spectrum diverges at the point $\lambda=\aal$, for $D=2$ it tends to a constant, and for $D\geq 3$ it vanishes. The thick lines show the exact spectrum~\ref{eq:spec}, while the soft dashed lines indicate the asymptotic predictions on the basis of Eq.~(\ref{eq:density_multidim}). \emph{(Right panel)} Average over components of the autocorrelation function for the same system at the critical value $\aal=1$.
While at $D\leq 2$ the correlation has no finite limit for $\aal\to 1$, we have represented its limiting value for $D=3,4,5$. While the heavy curves refer to the predictions of Eq.~(\ref{eq:spec}), the soft dashed lines plotted for comparison are the asymptotic results of Eq.~(\ref{eq:density_multidim}), predicting a decay of the type $\tau^{1-D/2}$.}
  \label{fig:correl_lattice}
\end{figure}

The spectrum $\rho(\lambda)$ for a system with next-neighbor interactions is represented in Fig.~\ref{fig:correl_lattice} together with the average over components of the autocorrelation function $\avcd{c(\tau)}$.
We remark that the above results are valid not only in the regime in which the large size limit $N\to\infty$ is taken before the $\aal\to 1$ one, but also when $\aal$ tends to the critical value $\aal = 1$ slower than $\aal = 1-K/N$ (see App.~\ref{app:fin_dim}). This scaling induces a maximum decay time for correlations, as it bounds the characteristic times of the various decay modes to be smaller than $N/K \beta$. Then, we expect the range of time over which the power-law decay of correlations is observed in a linear system with short range interactions to be at most of the order of the system size $N$.

\subsubsection{Case of a non-parabolic spectrum ($D=1$)}
\label{sec:crit_dim_1}
We want to stress here that a non-trivial behavior may also emerge as a consequence of the non-analyticity of the spectrum around its maximum. In fact, in the preceding section we have been able to argue that if the spectrum of a one-dimensional system $\lambda(\veck)$ is twice-differentiable, the $L\to\infty$, $\bar\alpha\to 1$ limit of the autocorrelation function is infinite (where we remind that in order not to obtain trivial results we are taking the $L\to\infty$ limit before the other ones). Suppose instead that the one-dimensional interaction matrix has tails proportional to $\alpha_\veci \sim |\veci|^{-1-\gamma}$, as for example in the case
\begin{equation}
  \label{eq:long_range_kern}
  \alpha_\veci = \bar \alpha \left( \frac {(1+|\veci| \bmod L)^{-\gamma-1} + (1-(L-|\veci|) \bmod L)^{-\gamma-1} }{2 \zeta(1+\gamma)} \right)\; ,
\end{equation}
where $\zeta(1+\gamma)$ is the Riemann Zeta function calculated in $1+\gamma$ for $\gamma>0$. Then the large $L$ limit of the spectrum can be calculated explicitly, and results
\begin{equation}
  \label{eq:spectrum_long_range_kernel}
	\lambda(\veck) = \bar \alpha (e^{ik}\mathcal{P}[1+\gamma,e^{ik}] + e^{-ik}\mathcal{P}[1+\gamma,e^{-ik}]) \; ,
\end{equation}
where $\mathcal{P}[1+\gamma,z] $ is the Polylogaritmic function of order $1+\gamma$ calculated in $z$.
Its maximum $\lambda(\vecz)$ is equal to $\bar\alpha$, but its first derivative at $\veck=0$ is not defined for $0<\gamma<1$. The analysis of the behavior of the spectrum in the vicinity of $\veck=0$ in that case reveals that the leading term in the expansion of $\lambda(\veck)$ around $\vecz$ is proportional to $||\veck||^\gamma$, and thus in the limit $\bar\alpha\to 1$ induces a long-time behavior of the autocorrelations of the type
\begin{equation}
  \label{eq:1}
\avcd{c(\tau)} \sim \tau^{-(1-\gamma)/\gamma} \;.
\end{equation}
Intuitively, by diluting the interactions among a larger number of components, it is possible to tame the strong fluctuations arising in one-dimensional systems, obtaining a finite limit for the correlations.
This also implies that long-range correlations are present for $1>\gamma>1/2$, when the dilution is less severe.

\section{The stochastic case}
While the definition of a periodic system such as the one analyzed above requires the existence of a notion of geometry in the space of the coordinates, we want now to focus our analysis on the disordered case in which no particular geometry is present in the system. This is typically a more realistic scenario for complex systems  in which the interactions are not thought to be organized according to a peculiar spatial structure. In order to account for this lack of regularity, we model the parameters of the system  as random variables extracted from a given \emph{statistical ensemble}.
In this type of framework, characterizing the behavior of a linear model requires understanding how the ensemble fluctuations in the parameters defining the model are inherited by the intensities $\Lambda$ and the cross-correlations $c(\tau)$.

The first type of ensemble that we are going to consider is a suitable one in order to model disordered realizations of a VAR model, while it cannot be used to model Hawkes processes, as it assigns negative weights to the entries of the interaction matrix $\alpha$ with finite probability. Indeed, a great deal of results can be proved rigorously in this framework. Moreover, many of the following results can be used in order to have an intuitive grasp about more general scenarios, in which an analytical solution is  not necessarily available.

\subsection{The Gaussian Orthogonal Ensemble}
\label{sec:goe}
\subsubsection{Definition}
We first consider a statistical ensemble in which the $\alp$ matrices are drawn from the Gaussian Orthogonal Ensemble (GOE), in which each $\alp$ is drawn according to a weight
\begin{equation}
\label{eq:GOE}
P_N(\alp) \propto \exp \left( \frac{N}{4\sal^2} {\textrm{tr}}\left[ \left(\alp - \one \frac \aal N\right)\left(\alp^T - \one \frac \aal N \right)\right] \right) \; ,
\end{equation}
where $\one$ is the $N$-dimensional matrix with all components equal to one.
This choice implies
\begin{eqnarray}
\av{\mal} &=& \frac{\aal}{N} \\
\var{\mal}  &=&
\left\{
\begin{array}{ccc}
2 \sal^2 /N &\textrm{if} & i = j \\
&&\\
\sal^2 / N  &\textrm{if} & i \neq j
\end{array}
\right.
\end{eqnarray}
Where we are using the symbol $\av{\dots}$ in order to denote averages with respect to the measure defined by $P_N(\alp)$ and $\var{\dots}$ in order to denote the variances with respect to that same measure.
The factors of $N$ appearing in the definition of the ensemble have been chosen in order for both the mean and the fluctuations of each of the $N$ quantities $\{ \sum_{j=1}^N  \mal \}_{i=1}^N$ not to depend explicitly on $N$.
Intuitively, this implies that if $\aal$ and $\sal^2$ are finite in the large $N$ limit, then the interaction strength on each of the $N$ components is also finite.
Moreover, it is possible to prove (see~\cite{Mehta:2004aa}) that the matrix $\alp$ admits almost surely the decomposition
\begin{equation}
\label{eq:decomp_goe}
\alp = \one \frac{\aal}{N} + O \lambda O^T \; ,
\end{equation}
with $OO^T=O^TO=\id$, which will be more useful than the decomposition~(\ref{eq:decomp}) formerly introduced, due to the better symmetry properties of the $O$ matrices.
As usual, $\lambda$ is a diagonal matrix of eigenvalues $\lambda = \{ \delta_{ab} \lambda_a \}_{ab=1}^N$, whose joint probability can be written as
\begin{equation}
p(\lambda_1,\dots , \lambda_N) \propto e^{-\frac N {4\sal^2} \sum_a \lambda_a^2} \prod_{a\neq b} |\lambda_a - \lambda_b| \; ,
\end{equation}
and $O$ is an orthogonal matrix sampled uniformly and independently of $\lambda$ on the $N$-dimensional Haar sphere~\cite{Mehta:2004aa}.

Once the statistical ensemble for the interaction matrix $\alp$ is fixed, it is necessary to prescribe a rule for the statistics of the vector quantities $\mu_i$, $\Lambda_i$ and $\Sigma_{ii}$ appearing in the model. In particular, we will assume the  $\mu_i$ to be independent and identically distributed variables, with mean $\av{\mu_i}=\amu$ and variance $\var{\mu_i}=\smu^2$. The parameters $\Lambda_i$ will be indirectly fixed by the relation $\Lambda =  \hat \Psi(0) \mu$.
Finally, we additionally need to specify the statistics for the $\Sigma_{ii}$. We will take them to be independent and identically distributed with mean
$\av{\Sigma_{ii}}=\asi$ and variance $\var{\Sigma_{ii}}=\ssi^2$ in the VAR case, while in the Hawkes model the values of $\Sigma$ will be indirectly fixed by the relation $\Sigma_{ii}=\Lambda_i$.

\subsubsection{Stability in the GOE ensemble}
The statistical ensembles defined above assigns strictly positive probability to $\alp$ matrices associated with unstable systems (i.e., with finite probability it can be that $||\alp|| >1$), spoiling the assumption \textbf{(H1)}. In order to focus our analysis to the stable realizations sampled from those ensembles, it is necessary to restrict the expectation $\av{\dots}$ to the set of matrices for which the largest eigenvalue is smaller than one. In particular, it will be useful to consider the probability measure $\overline{P_N}(\alp)$, defined as:
\begin{equation}
\label{eq:stable_cond}
\overline{P_N} (\alp) = P_N\Big(\alp \, \Big|\,   ||\alp|| < 1\Big) = \frac{P_N(\alp)}{P_N( || \alp || < 1)} \mathds{1}_{[0,1)}(|| \alp ||) \; .
\end{equation}
We denote averages taken with respect to this measure with $\avs{\dots}$. As we are interested in studying the stability of these linear models in the regime of large $N$, we will characterize the large $N$ behavior of  $P_N( || \alp || < 1)$, and in verifying whether $P_N( || \alp || < 1)$ tends to a non-vanishing constant $P_\infty( || \alp || < 1)$ in the limit of large $N$.

\begin{figure}
  \centering
  \includegraphics{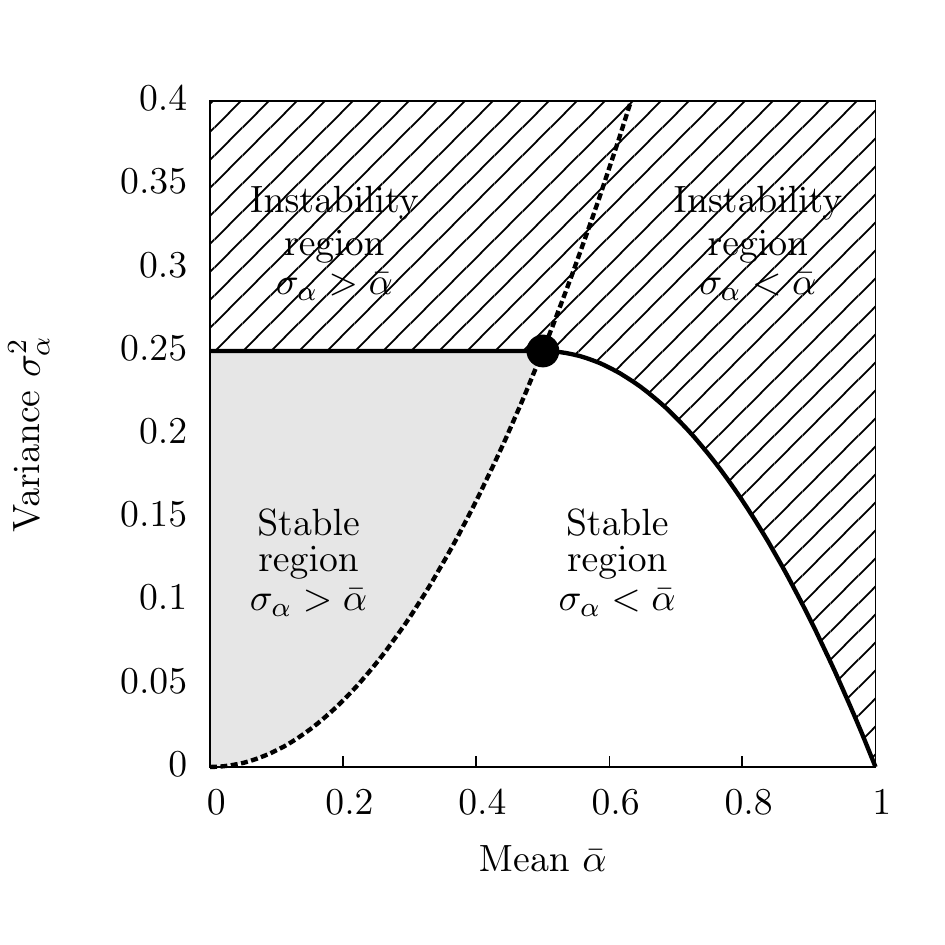}
  \caption{Phase space for the VAR model as a function of the parameters defining the statistical ensemble (inthe GOE ensemble) for the matrix $\alp$ (see Section \ref{sec:goe}). A region of stability (in which $P_\infty(||\alp || < 1)=1$) is separed from an unstable region (where $P_\infty(||\alp || < 1)=0$) by a critical line in which $P_\infty(||\alp || < 1)$ is finite. While for $\aal > \sal$ the maximum eigenvalue is isolated, along the $\aal < \sal$ portion of the critical line (in bold) the maximum eigenvalue corresponds to the edge of the support for the density of eigenvalues.}
  \label{fig:phase_diagram_goe}
\end{figure}

Indeed, the problem of stability in the GOE can be solved exhaustively by exploiting the results of Ref.~\cite{Furedi:1981aa,Feral:2007aa,Baik:2005aa} (also see Ref.~\cite{Arous:2011aa} for a review), where it is shown that the behavior of the largest eigenvalue of the matrix $\alp$ is dictated by the ratio between $\aal$ and $\sal$.
\begin{itemize}
\item
For $\aal / \sal > 1$, the largest eigenvalue $\lambda_{max}$ is a Gaussian variable of mean $\aal + \sal^2/\aal$ and variance $\sal^2 / N$.
\item
If $\aal / \sal < 1$, the distribution of the rescaled random variable $( \lambda_{max}/\sal - 2 ) N^{2/3} $ converges to the Tracy-Widom distribution \cite{Tracy:1994aa,Tracy:2000aa}.
\end{itemize}
This implies that, if $\aal / \sal > 1$ and $1-\aal - \sal^2/\aal \gg N^{-1/2} $, then drawing an unstable sample matrix $\alp$ becomes a large deviation event, and $P_\infty(||\alp || < 1) = 1$. In the opposite case, if $\aal / \sal < 1$, it is enough that $1-2\sal \gg N^{-2/3}$ in order to have $P_\infty(||\alp || < 1) = 1$ (for an exhaustive discussion accounting for this kind of large deviations see Ref.~\cite{Dean:2008aa,Dean:2006aa}). In the special case in which the maximum eigenvalue is close to the instability point ($\lambda_{max}=1+O(N^{-1/2})$ for the case of $\aal/\sal$ large or $\lambda_{max}=1+O(N^{-2/3})$ for $\sal/\aal$ small), then $P_N(||\alp || < 1)$ tends to a non-vanishing constant depending upon the precise values of $\aal$ and $\sal$.
This behavior is summarized in Fig.~\ref{fig:phase_diagram_goe}, where we show the shape of the stability region in the $(\aal,\sal^2)$ plane. Notice that this type of transition is precisely the one discovered in Ref.~\cite{May:1972aa}, which has been later related to the onset of a third order phase transition between the \emph{pulled} and \emph{pushed} phase of a Coulomb gas~\cite{Majumdar:2014aa}.\\
Additionally, an important qualitative difference emerges in the phases $\aal < \sal$ and $\aal > \sal$: while in the former case at large $N$ the largest eigenvalue coincides with the support of the spectral density of eigenvalues, in the latter there is a gap among the support of the eigenvalue distribution and the largest eigenvalue (see Sec.~\ref{sec:crit_goe} and Fig.~\ref{fig:spectrum_goe}). We will see that this difference will have a central role in determining the limiting behavior of correlations in the critical case.

\subsubsection{Endogeneity and Relaxation in the GOE ensemble}
\label{sec:obs_goe}
Having established the stability condition for the matrices in the GOE, we enunciate our result for the averages of $\avcd{\lambda}$, $\avcd{\hat c(\omega)}$ and $\avcm{\hat c(\omega)}$. The systematic procedure used to obtain these results is illustrated in App.~\ref{app:obs}. We get:
\begin{eqnarray}
\avs{\avcd{\Lambda}}
\label{eq:mean_lambda_goe}
&=&
\frac{\amu}{\aal} \, \left( \avs{\gf{1} (\aal,\{ (1-\lambda_a)^{-1}\})}-1 \right)
\\
\avs{\avcd{ \hat c(\omega)} }
&=&
\aal^2 \frac{\asi}{2N} ||\hat\phi(\omega)||^{-2} \partial_{x^{(3)}}^2 \avs{\gf{3} ((\aal,\aal,0),\{ z_a^\star(\omega)\},\{ z_a(\omega)\},\{ ||z_a(\omega) ||^2\}} \nonumber \\
&+& \aal \frac{\asi}{N} ||\hat\phi(\omega)||^{-2} \partial_{x^{(2)}} \avs{\gf{2}((\aal,0),\{ z_a^\star(\omega)\},\{ z_a^\star(\omega) ||z_a(\omega) ||^2 \}} \nonumber \\
&+& \aal \frac{\asi}{N} ||\hat\phi(\omega)||^{-2} \partial_{x^{(2)}} \avs{\gf{2} ((\aal,0),\{ z_a(\omega)\},\{ z_a(\omega) ||z_a(\omega) ||^2\}} \nonumber \\
&+& \frac{\asi}{N} ||\hat\phi(\omega)||^{-2} \avs{\sum_a \frac{1}{|| \hat \phi^{-1}(\omega)  -\lambda_a||^2}} \label{eq:mean_acorr_goe}
\\
\avs{\avcm{ \hat c(\omega)} }
&=&
\frac \asi N ||\hat\phi(\omega)||^{-2} \partial_{x^{(3)}} \avs{\gf{3} ((\aal,\aal,0),\{ z_a^\star(\omega)\},\{ z_a(\omega)\},\{ ||z_a(\omega) ||^2\}}\nonumber  \; , \\
& & \label{eq:mean_xcorr_goe}
\end{eqnarray}
with $z_a(\omega) = (\hat \phi^{-1}(\omega) -\lambda_a)^{-1}$. The generating function $\gf{p}(\vec x,\{z_a^{(k)}\}_{a=1}^N) $ appearing in above formulae can be defined after considering a $p$-dimensional vector $\vec x =(x^{(1)},\dots , x^{(p)})$ and a matrix $\{ z_a^{(k)}\}_{(a,k) \in (1,\dots , N) \times (1,\dots p)}$, in whose case it can be written as:
\begin{eqnarray}
\gf{p}(\vec x,\{z_a^{(k)}\}_{a=1}^N) &=& \frac{\Gamma(N/2)}{\Gamma(N/2-p)} \int_0^1 dt_1 \dots dt_p \prod_{k=1}^p t_k^{k-1} (1-t_p)^{N/2-p-1} \nonumber \\
&\times & \prod_a \left(1 - \sum_{k=1}^p h^{(k)}(\vec t) x^{(k)} z_a^{(k)} \right)^{-1/2} \;,
\end{eqnarray}
together with a set of auxiliary functions $h^{(k)}(\vec t)$ given by
\begin{eqnarray}
h^{(1)} &=& t_1 \dots t_p \\
h^{(2)} &=& (1-t_1) t_2 \dots t_p \\
h^{(3)} &=& (1-t_2) t_3 \dots t_p \\
&\dots&\\
h^{(p)} &=& (1-t_{p-1}) t_p  \; .
\end{eqnarray}
Eqs.~(\ref{eq:mean_lambda_goe}),~(\ref{eq:mean_acorr_goe}) and~(\ref{eq:mean_xcorr_goe}) above allow to express the observables as averages over the eigenvalue distribution of the  generating functions $\gf{p}(\vec x,\{z_a^{(k)}\}_{a=1}^N)$. In App.~\ref{app:obs} we illustrate how to express a generic observable within the model in terms of these generating functions, and show how to use this formalism in order to derive a systematic expansion in powers of $\aal$ for the momenta of any observable. For example, in the case of the mean intensities, the first term of such an expansion, corresponding to the case $\aal=0$, reads:
\begin{eqnarray}
\avs{\avcd{\Lambda}}
&=&
\frac{\amu}{N} \avs{\sum_a \frac 1 {1-\lambda_a}}
\label{eq:mean_lambda_goe_al0}
\\
\avs{\avcd{ \hat c(\omega)} }
&=&
\frac{\asi}{N} ||\hat\phi(\omega)||^{-2} \avs{\sum_a \frac{1}{|| \hat \phi^{-1}(\omega) -\lambda_a ||^2 }}
\\
\avs{\avcm{ \hat c(\omega)}}
&=&
\frac{\asi}{N^2} ||\hat\phi(\omega)||^{-2} \avs{\sum_a \frac{1}{|| \hat \phi^{-1}(\omega) -\lambda_a ||^2 }} \; .
\end{eqnarray}
Even more interestingly, the leading order term of the  $1/N$ expansion of $\gf{p}$ can be computed analytically (see App.~\ref{app:obs}), making possible the asymptotic estimation of the observables by means of the $\gf{p}$ generating functions. \\

Eqs.~(\ref{eq:mean_lambda_goe}) and~(\ref{eq:mean_lambda_goe_al0}) state that, unlike in Eq.~(\ref{eq:mean_lambda_lattice}), all modes of $\alpha$ contribute to the average intensity due to the heterogeneity of the interaction network. As in that case, if any of the eigenvalues exceeds the value $\lambda=1$, the system loses its stability, and the ratio of endogenous-to-exogenous intensity $\avcd{\Lambda}/\avcd{\mu} -1$ explodes.
The equations for the correlations for $\aal=0$ are similar to the ones found in the deterministic case, except for the fact that the average cross-correlation for the non-diagonal terms $\hat c_{ij}(\omega)$ with $i\neq j$ is exactly equal to zero.

We remark moreover that by specializing the value of $\avs{\hat c_{ii}(\omega)}$ to $\omega =0$, we find an expression proportional to the one of $\avs{\Lambda_{i}^2}$. This implies that the \emph{memory} $\hat c_{ii}(0) = \int d\tau \, c_{ii}(\tau)$ is related to the fluctuations of the mean. In particular, whenever the process develops long-range memory (i.e., $\hat c_{ii}(0)$ diverges), then the fluctuations of the mean intensities are also bound to diverge.

Finally, we remind that in the cases in which $P_\infty(||\alp||<1)$ is equal to one, then $P_\infty(\alp \, | \, ||\alp||<1) = P_N(\alp) \mathds{1}_{[0,1)}(|| \alp ||)$, so that it is possible to replace the unconditional measure with the actual one in the large $N$ limit. This is the case in all the stability region $\aal+\sal^2/\aal <1$, implying that, for example, the $\aal=0$ value of the average auto-correlation function in the exponential kernel case is given by
\begin{equation}
\avs{\avcd{  c(\tau) }}
\label{eq:acorr_goe_wigner}
=
\beta \asi \int_{-2\sal}^{+2\sal} d\lambda \, \rho_W(\lambda)  \frac{\lambda (2-\lambda) }{2(1-\lambda )} e^{-\beta [1-\lambda] |\tau|} \; ,
\end{equation}
where
\begin{equation}
\rho_W(\lambda) = \av{\frac{1}{N} \sum_a \delta(\lambda - \lambda_a)}= \frac{1}{2 \pi \sal^2} \sqrt{4\sal^2-\lambda^2}
\end{equation}
is the well-known Wigner semicircle law with support in $[-2\sal,2\sal]$ represented in Fig.~\ref{fig:spectrum_goe} (see \cite{Wigner:1955aa}). Fig.~\ref{fig:correl_goe} shows a set of auto-correlation curves obtained by varying $\sal$ along the $\aal=0$ line, comparing the results of Eq.~(\ref{eq:acorr_goe_wigner}) with the results of numerical simulations.

\subsubsection{Fluctuations and law of large numbers for $\avcd{\Lambda}$}
A natural question arising in the study of systems with random interactions is the one related to the self-averagingness of the observables: Given a succession of realizations of the system of increasing $N$, does a law of large numbers hold for the quantities $\avcd{\Lambda}$ and $\avcd{ c(\tau)}$?
Indeed the mean intensity $\avcd{\Lambda}$ depends upon the eigenvalues $\{ \lambda_a\}_{a=1}^N$, which are strongly correlated variables. As a consequence, one may expect the natural scaling of fluctuations to be altered by the particular statistics of the eigenvalues. The fluctuations of the mean intensity $\avcd{\Lambda}$ can be computed analytically by using the results of App.~\ref{app:obs}, and result
\begin{eqnarray}
\avs{\avcd{\Lambda\Lambda^T}}
&=&
\amu^2\left( \aal \partial_{x^{(1)}}  + 1 \right) \partial_{x^{(2)}} \avs{\gf{2} ((\aal,0),\{ (1-\lambda_a)^{-1}\},\{ (1-\lambda_a)^{-2}\})} \nonumber \\
&+&  \aal^2 \frac{\sigma^2_\mu}{2N} (\aal \partial_{x^{(1)}} + 1) \partial_{x^{(2)}}^2 \avs{\gf{2} ((\aal,0),\{ (1-\lambda_a)^{-1}\},\{ (1-\lambda_a)^{-2}\})} \nonumber \\
&+&  \aal \frac{2 \sigma^2_\mu} {N} \partial_{x^{(2)}} \avs{\gf{2} ((\aal,0),\{ (1-\lambda_a)^{-1}\},\{ (1-\lambda_a)^{-3}\})}  \nonumber \\
&+& \frac{ \sigma^2_\mu} N \avs{\sum_a \frac 1 {(1-\lambda_a)^2}}
\label{eq:fluct_diag_lambda_goe}
\\
\avs{\avcm{ \Lambda\Lambda^T}}
&=&
\frac{\amu^2}{\aal^2} (\aal \partial_{x^{(1)}} - 1) \avs{\gf{1} (\aal,\{ (1-\lambda_a)^{-1}\})   - 1 -\frac{\aal}{N}\sum_a \frac 1 {(1-\lambda_a)} } \nonumber  \\
&+&  \frac{\sigma^2_\mu}{N} (\aal \partial_{x^{(1)}} + 1) \partial_{x^{(2)}} \avs{\gf{2} ((\aal,0),\{ (1-\lambda_a)^{-1}\},\{ (1-\lambda_a)^{-2}\})} \; , \nonumber \\
&& \label{eq:fluct_coll_lambda_goe}
\end{eqnarray}
While Eq.~(\ref{eq:fluct_diag_lambda_goe}) is useful to estimate the fluctuations of the individual $\Lambda_i$, Eq.~(\ref{eq:fluct_coll_lambda_goe}) can be used in order to estimate the variance of $\avcd{\Lambda}=N^{-1}\sum_i\Lambda_i$.

\vskip .3cm
\noindent
{\bf Considering the case $\aal=0,\sal < 1/2$.}
In particular, we are able to estimate analytically the latter quantity along the $\aal=0,\sal < 1/2$ line, as in such region the mean intensity reduces to a linear statistics of the eigenvalues. In that case, for $N$ large enough the variance of $\avcd{\Lambda}$ tends to
\begin{eqnarray}
\label{eq:lin_stat}
\vas{ \avcd{\Lambda} } &\to&
\frac{\smu^2}{N} \int_{-2\sal}^{2\sal} d\lambda  \frac {\rho_W(\lambda)} {(1-\lambda)^2}  \\
&+&\frac{\amu^2}{N^2}
 \mathbb{P} \left\{ \int_{-2\sal}^{2\sal} d\lambda d\lambda^\prime \frac{\rho_W(\lambda,\lambda^\prime)}{(1-\lambda)(1-\lambda^\prime)}  \right\}\; , \nonumber
\end{eqnarray}
where $\mathbb{P}$ denotes the Cauchy principal value, and
\begin{eqnarray}
  \label{eq:spectral_corr}
  \rho_W(\lambda,\lambda^\prime) &=& -\frac 1 {\pi^2} \left( \frac{1}{[(2\sal+\lambda)(2\sal-\lambda)]^{1/2}} \right) \\
&\times& \frac{\partial^2}{\partial \lambda \partial \lambda^\prime} \left([(2\sal+\lambda^\prime)(2\sal-\lambda^\prime)]^{1/2} \log |\lambda-\lambda^\prime| \right)
\end{eqnarray}
is an universal \emph{spectral correlation}, which encodes the strongly interacting nature of the eigenvalue distribution~\cite{Brezin:1993aa,Beenakker:1994aa}. As both integrals appearing in Eq.(\ref{eq:lin_stat}) converge, the contribution of the $\smu$ and $\amu$ terms scale respectively as $N^{-1}$ and $N^{-2}$.

\vskip .3cm
Let us point out that we find by numerically evaluating Eq.~(\ref{eq:fluct_coll_lambda_goe}) that this scaling extends to all the values of $(\aal,\sal)$ in the interior of the stability region.
The scaling of fluctuations in the critical regime is
indeed non-trivial: taking for example the case $\aal=0$, one can check that both integrals appearing in Eq.~(\ref{eq:lin_stat}) are formally divergent, indicating the emergence of a different scaling with $N$ of the fluctuations. Finding the exact scaling at transition requires indeed a more sophisitcated analysis, as one needs to take into account that $P_N(\alp)\neq \overline{P_N}(\alp)$ in order to compute the fluctuations of the largest eigenvalues dominating the divergence.

\subsubsection{Criticality}
\label{sec:crit_goe}
\begin{figure}
  \centering
  \includegraphics{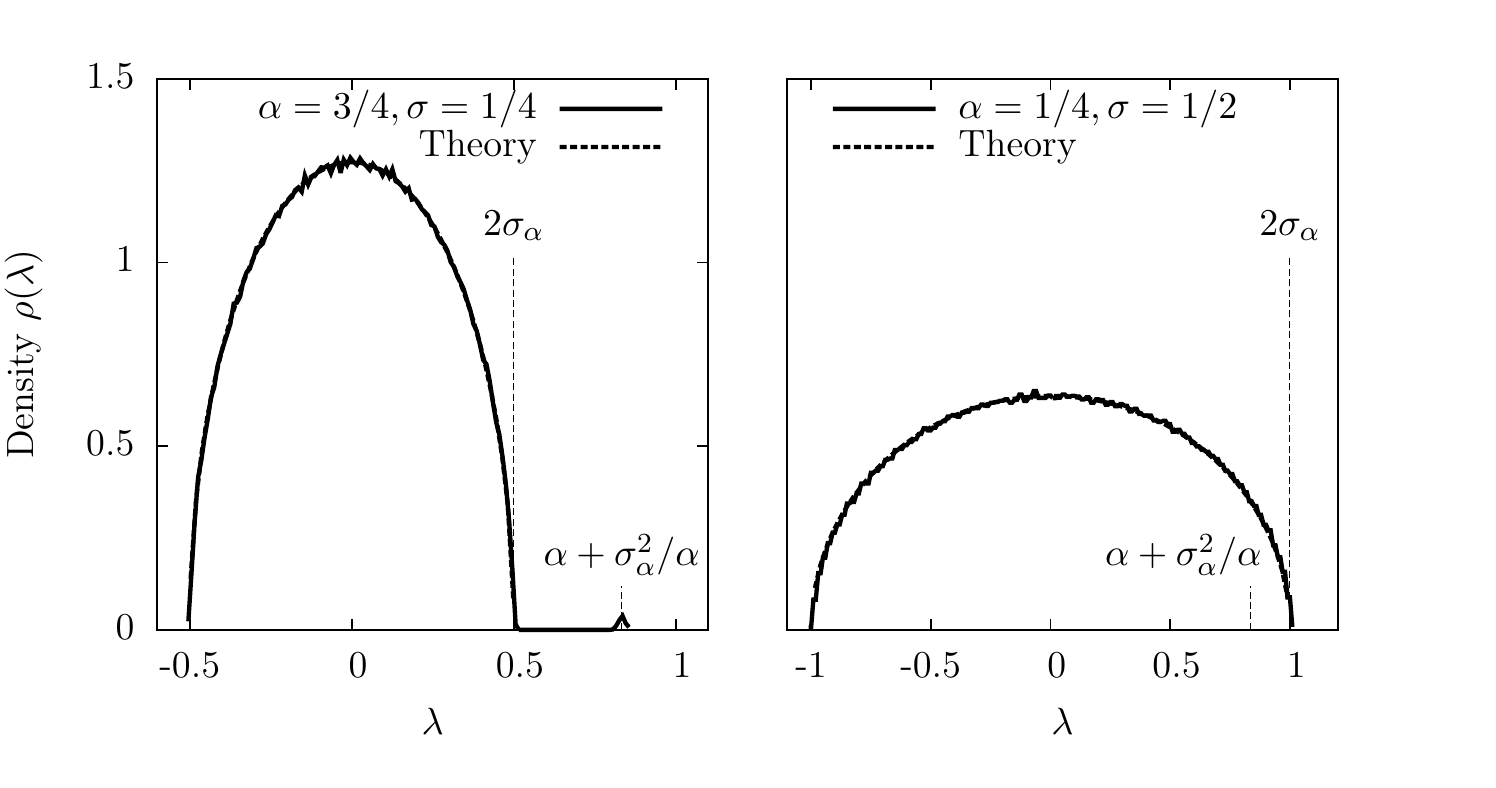}
  \caption{Spectrum $\rho(\lambda)$ of the Gaussian Orthogonal Ensemble in the cases $\aal>\sal$ (left panel) and $\aal < \sal$ (right panel) (See Section \ref{sec:goe}). We have averaged over 100 realizations of a system of size $N=1000$ in order to obtain the curves depicted in the figure, which at this scale are almost exactly superposed to the Wigner semicircle.}
  \label{fig:spectrum_goe}
\end{figure}

Similar to what has been discussed above for a translationally invariant system, even in this random framework we can imagine the system to be able to develop slow correlation if the spectrum touches the instability point $\lambda=1$. The situation is indeed more delicate in this case, as the maximum value may not to coincide with the edge of the support of $\rho(\lambda)$ as it was in the former case.
In particular when $\aal > \sal$ and $\alpha+\sal^2/\alpha=1$ the large time behavior is dictated by an isolated exponential mode whose associated decay speed becomes extremely small. This scenario is similar to the one considered in Ref.~\cite{Bremaud:2001aa}. The phase in which $\aal < \sal$ and $\sal$ is close to the instability point $\sal=1/2$ leads instead to a richer dynamical behavior, and its phenomenology is, to the best of our knowledge yet unexplored.
Fig.~\ref{fig:spectrum_goe} summarizes this description by representing the spectrum in both of these cases.

\begin{figure}
  \centering
  \includegraphics{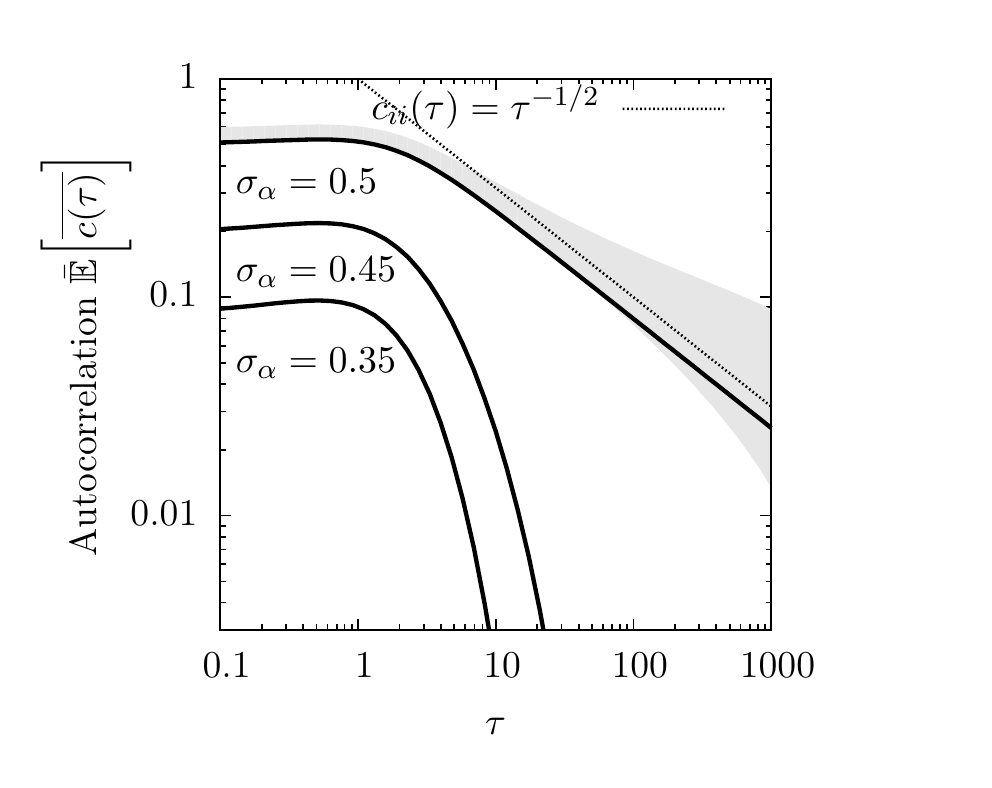}
  \caption{Autocorrelation function of the GOE ensemble for variable $\sal$ in the case $\aal=0$ (See Section \ref{sec:obs_goe}). The solid lines represent the theoretical predictions, while the shaded regions are two-sigmas error bars accounting for the results of the numerical estimations (300 realizations of systems of size $N=1000$). Notice that the theoretical predictions for $\sal<1/2$ are indistinguishable from the numerical results, while for $\sal=1/2$ strong fluctuations arise in the system as a consequence of the critical regime that we are considering.}
  \label{fig:correl_goe}
\end{figure}

\vskip .3cm
\noindent
{\bf Considering the case $\aal=0,\sal = 1/2$.}
In order to be definite, let us consider the case $\aal=0$ and $\sal =1/2$. In that case and in the large $N$ regime the realizations are stable with finite probability $P_\infty(||\alp|| < 1) = \mathcal{P}_{1}(0) \approx 0.8319$, where $\mathcal{P}_{\beta}(z)$ is the cumulative of the Tracy Widom distribution of index $\beta$ calculated in $z$ (see Refs.~\cite{Dean:2008aa,Dean:2006aa}).
 The long time regime of the correlations is then dictated by the behavior of the density of eigenvalue close to 1. In particular, if one considers
 \begin{equation}
 \bar \rho_W(\lambda) = \avs{\frac{1}{N} \sum_a \delta_{\lambda - \lambda_a}} \;,
 \end{equation}
the behavior of $ \avs{\avcd{  c(\tau) }}$ depends upon the shape of $ \bar \rho_W(\lambda) $ close to 1. Notice that, as $P_\infty(||\alp||<1) < 1$, one could expect that $ \bar \rho_W(\lambda)  \neq \rho_W(\lambda)$. However, at leading order in $N$, the two densities converge to the same one, and it is possible to use Eq.~(\ref{eq:acorr_goe_wigner}) even in the critical regime \cite{Dean:2008aa,Dean:2006aa}. Intuitively, this indicates that the number of eigenvalues exceeding the instability point $\lambda=1$ is of order smaller than $O(N)$, implying that they give a negligible contribution to $\bar \rho_W(\lambda)$ in the large $N$ limit. In the case of the exponential kernel, we find by plugging $\bar \rho_W(\lambda)$ into Eq.~(\ref{eq:acorr_factor_real}) that for $N$ large enough
\begin{equation}
  \label{eq:ac_crit_wigner}
  \avs{ \avcd{ c(\tau)}} = \frac{\asi \beta} \pi \int_{-1}^{1}d\lambda \, \lambda (2-\lambda)
\sqrt{ \frac{1+\lambda}{1-\lambda}}e^{-\beta|\tau|(1-\lambda)}
\sim \tau^{-1/2} \; .
\end{equation}
Studying the tail behavior of the average autocorrelation is indeed  delicate: one can rigorously finds a scaling $c_{ii}(\tau) \sim \tau^{-1/2}$ just in the setting in which the limit of large $N$ is taken before the one of large $\tau$. In practice the finite $N$ corrections establish a upper cutoff to the correlations, which decay exponentially beyond a time $\tau_{cut}$ whose location depends upon the corrections to the limiting law of $ \bar \rho_W(\lambda) $. Moreover, fluctuations are large at any $\tau$: while for $\sal<1/2$ the variance of $\sum_i  c_{ii}(\tau) $ has a  finite limit for $N\to\infty$, in the case $\sal=1/2$ such limit becomes infinite. The behavior of the autocorrelations in the critical case $\aal=0,\sal=1/2$ is also represented in Fig.~\ref{fig:correl_goe}, where we have compared the theoretical predictions with the results of numerical simulations.
The divergence of $\hat c_{ii}(0) = \int d\tau\, c_{ii}(\tau)$ in this regime signals that long-range memory is induced in the system as an effect of the structure of the interaction matrix $\alp$.
Notice that the behavior of $\hat c_{ii}(\omega)$ in $\omega=0$ and the one of the fluctuations of the mean intensities $\Lambda_i$ are related. In particular, if $\hat c_{ii}(0)$ diverges, Eq.~(\ref{eq:fluct_diag_lambda_goe}) implies that the mean intensity $\Lambda_i$ has infinite variance.

\vskip .3cm
We remark that, as in the translationally-invariant case discussed above, the slow relaxation of the correlations is independent on the specific shape of the kernel $\phi(\tau)$, as long as the integral $\int_0^\infty d\tau\,\tau \phi(\tau)$ is finite.

\subsection{Random  graph ensembles}
\label{sec:rrg}
\subsubsection{Definition}
In this section we consider ensembles in which the interaction matrix $\alp$ is generated by randomly assigning positive weights to the edges of a graph whose vertices are the $N$ coordinates of the system. As in this setting the matrix $\alp$ is by construction non-negative, we are free to use the following ensembles in order to build random realizations of a multivariate Hawkes proces.\\
There is indeed an important difference to underline among VAR processes and multivatiate Hawkes models. In fact, while the $\Sigma_{ii}$ appearing in a VAR model are independent of the $\Lambda_i$ variables, in the Hawkes model their values need to coincide. This implies in particular that the expression of the correlations in the two cases can in general be different.
Summarizing, while at large times a deterministic Hawkes process is always identical to an appropriately tuned continuous-time VAR, in the random case they can be different, as the noise in the Hawkes case is endogeneously generated.

As the ensemble for non-negative realizations of $\alp$, we take the one of the \emph{$c$-regular random graphs}, in which each $\alp$ is associated with an adjacency matrix $\mathcal{G}$ uniformly chosen in the space of random undirected graphs of $N$ vertices, each of which is connected exactly to $c$ vertices. We then define $\alp$ as $\alp=(\aal / c) \mathcal G$. According to these conventions, the first two momenta of each entry result
\begin{eqnarray}
\av{\mal} &=& \frac{\aal}{N} \\
\var{\mal}  &=& \frac{\aal^2}{cN} \left(1- \frac c N \right) \; .
\end{eqnarray}
We will further suppose $c$ to be an $N$-independent constant. In fact, if $c$ grows with $N$, the bulk of eigenvalues of $\alpha$ shrinks with $N$, leading in the large $N$ limit to an asymptotic spectrum consisting of $N-1$ degenerate eigenvalues equal to zero and to a unique eigenvalue equal to $\aal$ \cite{Kesten:1959aa,McKay:1981aa}.
This type of behavior has been proved for several other ensembles of non-negative $\alpha$ matrices (e.g, \cite{Erdos:2012aa,Erdos:2013aa,Bordenave:2011aa}), and can be informally summarized by the statement that \emph{ensembles of non-negative $\alpha$ matrices are expected to have a non-degenerate behavior in the large $N$ limit just in the sparse case}. By \emph{sparse}, we mean that given a random row $1\leq i\leq N$, a number of entries independent of $N$ should account for a finite fraction of the sum $\sum_j \mal$ with large probability.

\subsubsection{Stability in the regular random graph ensemble}
The issue of stability in this ensemble is trivial, as for $\aal<1$ the system is always stable (the maximum eigenvalue does not fluctuate because the degree of each node of the graph is fixed), and the measure associated with the averages $\avs{\dots}$ coincides with the unrestricted one $\av{\dots}$. Indeed, the maximum eigenvalue of the matrix $\alpha$ is associated with the eigenvector $U_{ia}=N^{-1/2} (1,\dots,1)$. This scenario is analogous to the one analyzed in Sec.~\ref{sec:det}, in which the absence of fluctuations for the maximum eigenvalue led the Perron-Froebenius eigenvalue to be associated with the same eigenvector.

\subsubsection{Endogeneity and Relaxation in the regular random graph ensemble}
\label{sec:observ-regul-rand}
The expressions for the averages of the various observables in the random regular graph are similar to the ones calculated in Sec.~(\ref{sec:det}) due to the absence of fluctuations in the largest eigenvalue of $\alpha$. By using the decomoposition~(\ref{eq:decomp}) we can in fact show that
\begin{eqnarray}
\avs{\avcd{\Lambda}}
&=&
\frac{\amu}{1-\aal}
\\
  \label{eq:ac_rg}
  \avs{\avcd{ \hat c(\omega)}} &=& \frac {\asi} N ||\hat \phi(\omega)||^{-2} \avs{\sum_{a} \frac 1 {||\hat \phi(\omega)^{-1} - \lambda^{a}||^2 }} \\
  \avs{\avcm{\hat c(\omega)}} &=& \frac {\asi} N ||\hat \phi(\omega)||^{-2}\frac 1 {||\hat \phi(\omega)^{-1} - \aal||^2 } \; .
\end{eqnarray}
\\

These expressions are valid for both the VAR and the Hawkes model, once one identifies $\Sigma_{ii}$ with $\Lambda_i$. This is because even though the correlations should be calculated differently in each of the two cases, the absence of fluctuations for the maximum eigenvalue induces the same expression of the correlations.\footnote{Including degree fluctuations in the graph associated to $\alpha$ would lead to a different expression of the correlations. In particular, the value of auto- and cross-correlations in the Hawkes case would be sensitive to the \emph{third} power of $\hat \Psi(\omega)$, as opposed to the \emph{second} power of $\hat \Psi(\omega)$ which appears in the expressions which we have reported.}
In the $N\to\infty$ limit, the form of the spectrum $\rho(\lambda)$ for this ensemble has been calculated in Refs.~\cite{Kesten:1959aa,McKay:1981aa}.
It results:
\begin{equation}
\label{eq:spectrum_rg}
\rho_c(\lambda) = \av{\frac{1}{N} \sum_a \delta_{\lambda - \lambda_a}}= \frac{1}{2 \pi \aal^2} \frac{\sqrt{4 \aal^2(c-1)-c^2\lambda^2}}{\aal^2-\lambda^2} \;,
\end{equation}
with support  $\lambda\in [-2\aal\sqrt{c-1}/c,2\aal\sqrt{c-1}/c]$. This explicit form of the spectral density can be used in order to evaluate numerically the diagonal terms $\avcd{\Lambda\Lambda^T}$ and $\avcd{\hat c(\omega)}$.

\subsubsection{Criticality}
\label{sec:crit_rrg}
The behavior of the system in the vicinity of the point $\aal=1$ can be analyzed easily due to the explicit expressions provided for the shape of the bulk (Eq.~(\ref{eq:spectrum_rg})) and the value of the maximum eigenvalue, which is always equal to $\aal$. As in the previous cases, we will refer explicitly to the case in which the kernel $\phi(\tau)$ is  an exponential function ($ \phi(\tau) = \beta e^{-\beta \tau}\mathds{1}_{\mathbb{R}^+}(\tau)$), reminding that this choice doesn't induce any loss of generality at large times as long as the kernel is assumed to be short-ranged.\\
For $c>2$ the edge of the bulk never touches the point $\aal$, indicating that the leading term of the large-time expansion of the auto-correlations is always proportional to $e^{-\beta (1-\aal)|\tau| }$. Then the $\aal\to 1 $ limit has the same type of behavior of the $\aal > \sal$ phase of the GOE ensemble already discussed. For $c=2$, the edge of the spectrum touches the point $\lambda=\aal$, as the density takes the form
\begin{equation}
\rho_{c=2}(\lambda) = \frac{1}{\pi \aal^2} ( \aal^2-\lambda^2)^{-1/2} \;.
\end{equation}
By inserting the above density into Eq.~(\ref{eq:ac_rg}) one finds the type of divergence already encountered in the case of a one-dimensional system: in order to have a finite limit for $\aal=1$, the leading behavior close to that point should be of the type $(1-\lambda)^\gamma$ with $\gamma>0$. This indicates that for $c=2$ in the limit $\aal\to 1$, the correlations are bound to become infinite.

It would nevertheless be interesting to find an ensemble of non-negative matrices in which (i) the fluctuations of the maximum eigenvalue tend to zero in the large $N$ limit,
(ii) the maximum eigenvalue touches the bulk of the spectrum and (iii) the spectral density vanishes at the edge of the spectrum. This would allow the model to describe the onset of non-trivial correlations at the critical point $\aal=1$ even for Hawkes processes. A natural candidate to this purpose would be the \er ensemble of $\alp$ matrices (informally, it can be thought of as an analytical continuation of the regular random graph to non-integer, fluctuating $c$). Moreover, the spectral density of the matrices sampled in that ensemble has been shown to be vanishing close to the edge of the bulk as in the Wigner case \cite{Bray:1988aa,Rodgers:1988aa,Biroli:1999aa,Semerjian:2002aa,Dorogovtsev:2003aa,Kuhn:2008aa}, allowing in principle non-exponential correlations to arise in the critical case. The problem of the \er ensemble lies indeed in points (i) and (ii), as the degrees of the nodes of the matrix $\alp$ can have potentially large fluctuations, leading to a slow divergence of the maximum eigenvalue in the large $N$ limit~\cite{Krivelevich:2003aa}, spoiling the stablility of the ensemble. Is it possible to generalize the \er ensemble by reducing the amount of degree fluctuations, so to control the maxmimal eigenvalue of $\alpha$? We leave this issue as an interesting open question.

\section{Conclusions}
\label{sec:conclusions}
In this work we have investigated the first and second order properties of two linear models (VAR and Hawkes processes) in the regime of high dimensionality, providing the values of the observables characterizing their behavior under different setups. We have addressed in particular the issues of stability, endogeneity, and slow relaxation for both deterministic and stochastic realizations of these models, showing that under the assumptions of factorizability {\bf (H2)} and homogeneity {\bf (H3)}, all the information about these systems can be related to the spectrum $\{\lambda_a \}_{a=1}^N$ of the matrix encoding its interaction network. 
We have shown that, as opposed to the univariate setup considered in~\cite{Bremaud:2001aa}, collective effects can trigger slow relaxation of the correlation functions even in systems with short-memory interactions provided that the edge of such spectrum \emph{smoothly} touches the instability point $\lambda=1$.
More generally, we find that in systems characterized by a strong degree of endogeneity, the dynamic properties are also expected to slow down, akin to what is observed for glassy systems~\cite{Kurchan:1996aa}, for which the proximity to a static transition point is signaled by a slow relaxation of the correlations.
We additionally relate the memory of the system with the self-averaging behavior of the intensity $\Lambda$: in the stochastic case long-range correlations are measured just when the variance of the observed signal diverges on a realization-per-realization basis. On a more general footing, we find that random realizations of a large linear systems are always self-averaging unless the system is exactly poised at the critical point.

Our study also illustrates the tight relation among the VAR and the Hawkes processes, whose first- and second-order properties are shown to coincide. A Hawkes process is in fact reminiscent of a VAR model in which the noise term is endogenously generated by the system. Such a fundamental difference also explain why random realizations of a Hawkes process differ from the ones sampled from a VAR, as in the former case the endogenous nature of the noise may enhance the level of fluctuations of the system, while in the latter the noise can be taken as completely exogenous.
Another fundamental difference among the two models is the excitatory nature of the interactions for a Hawkes system as opposed to a VAR model, in which the sign of interactions is arbitrary: Our study supports the hypothesis that a given degree of frustration (i.e., negative interactions) is required in order to observe slow relaxations in short-memory linear systems.

Finally, from the perspective of empirical calibration, our study shows that the parametric fit of a system characterized by power law relaxations with a short range kernel naturally leads to a critical point, the only one which can accommodate such a slow behavior of correlations in data. 

\section*{Acknowledgements}
We warmly thank F.~Lillo for long and fruitful discussions all along the preparation of this manuscript. We also thank J.-P.~Bouchaud, J.~Bun ,S.~Majumdar, P.~Vivo for useful discussions. This research benefited from the support of the ``Chair Markets in Transition'', under the aegis of ``Louis Bachelier Finance and Sustainable Growth'' laboratory, a joint initiative of \'Ecole Polytechnique, Universit\'e d'\'Evry Val d'Essonne and F\'ed\'eration Bancaire Fran\c{c}aise.

\appendix

\section{Spectral properties of a finite-dimensional model}
\label{app:fin_dim}
In Sec.~\ref{sec:det} we have analyzed the behavior of a linear model whose components lie on the vertices of a hypercube of dimension $D$. In this appendix we show how to calculate the $N\to\infty$ limit of the spectral density
\begin{equation}
  \label{eq:app_spec}
\rho(\lambda) = \frac 1 N \sum_{\veca} \delta(\lambda - \lambda_{\veca}) \;,
\end{equation}
for a model with a next-neighbor interactions matrix $\alpha_{\veci - \vecj}$. We will further match our result with the general prediction valid in the vicinity of $\lambda=\lambda_\vecz$ for any type of component-wise positive semi-definite $\alpha_{\veci - \vecj}$.
\subsection{Spectral density of a next-neighbor model}
The next-neighbor interaction kernel is defined by
\begin{equation}
  \label{eq:app_nn}
 \alpha_{\veci}=\frac{\aal}{2D} \sum_{d=1}^D \left( \delta_{\veci - \mathbf{e}_d}+\delta_{\veci + \mathbf{e}_d}\right)
\end{equation}
with $\mathbf{e}_d$ denoting the $d$-th element of the canonical basis of $\mathbb{R}^D$. By using Eq.~(\ref{eq:periodic_decomp}), it is straightforward to prove that
\begin{equation}
  \label{eq:app_spectrum}
  \lambda(\veck) = \frac{\aal}{D} \sum_{d=1}^D \cos\left( k_d \right)
\end{equation}
with $\veck = 2\pi \veca / L$. The limit $N\to\infty$ can be calculatied by writing the number of eigenvalues with density smaller than $\lambda$, which results
\begin{eqnarray}
  \label{eq:app_cumul_lambda}
  \mathcal{N}_<(\lambda)&=&\frac 1 N \sum_{\veca \in \{0,\dots,L-1\}^D} \mathds{1}_{\lambda_\veca \in (-\infty,\lambda)} \\
&\xrightarrow[N\to\infty]{}& \frac{1}{(2\pi)^D} \int_{[-\pi,\pi]^D} d\veck \, \mathds{1}_{\lambda(\veck) \in (-\infty,\lambda)} \nonumber \; .
\end{eqnarray}
By differentiation and successive exponentiation of a delta function, one finds
\begin{eqnarray}
  \label{eq:3}
  \rho(\lambda) &=& \frac{d\mathcal{N}_<(\lambda)}{d\lambda} = \frac{1}{(2\pi)^D}
\int_{-\infty}^{+\infty} \frac{dz}{2\pi}\int_{[-\pi,\pi]^D} d\veck \, e^{i z (\lambda - \frac{\aal}{D} \sum_{d=1}^D \cos\left( k_d \right))} \\
&=&\int_{-\infty}^{+\infty} \frac{dz}{2\pi} e^{i z \lambda} \left(\frac 1 {2\pi} \int_{-\pi}^{\pi} dk \, e^{-i z\frac{\aal}{D} \cos\left( k \right))}\right)^D \\
&=& \int_{-\infty}^{+\infty} \frac{dz}{2\pi} e^{i z \lambda} J_0^D(|z|\aal/D) \; ,
\end{eqnarray}
which, after a change of variables, is the result displayed in Eq.~(\ref{eq:spec}).

\subsection{Edge density for a poisitive definite interaction matrix}
We now want to compute the spectral density for a generic interaction matrix of the form $\alpha_{\veci - \vecj}$  which is positive semi-defninite and whose limiting spectrum is twice-differentiable in the limit $N\to\infty$.
In this case, the maximum of the spectrum is achieved for $\veck=\vecz$, so that one can expand it as
\begin{equation}
  \label{eq:expand_spectrum}
  \lambda(\veck)=\lambda(\vecz) + \frac{1}{2} \left(\veck^T H[\lambda(\vecz)] \veck \right) + o(\veck^2)
\end{equation}
Under this approximation, one can write that the number of eigenvalues larger than $\lambda$ is given by
\begin{eqnarray}
  \label{eq:expand_cumul}
  \mathcal{N}_>(\lambda) &\xrightarrow[N\to\infty]{}& \frac{1}{(2\pi)^D} \int_{\veck \in [-\pi,\pi]^D} d\veck \, \mathds{1}_{\lambda_\veca \in (\lambda,\lambda_{\vecz}]} \\
&\approx& \frac{1}{(2\pi)^D} \int_{\veck \in \mathbb{R}^D} d\veck \,  \mathds{1}_{-1/2 (\veck^T H \veck)\,  \in \,  [0,\lambda_{\vecz}-\lambda)} \;.
\end{eqnarray}
Above expression is the volume of a $D$-dimensional ellypsis, and it results
\begin{equation}
  \label{eq:approx_cumul}
   \mathcal{N}_>(\lambda(\vecz)-\epsilon) \approx \frac{\det^{-1/2}(-H[\lambda(\vecz)])}{(2\pi)^D} \frac{(2\pi\epsilon)^{D/2}}{\Gamma(D/2+1)}  \;.
\end{equation}
By differentiation with respect to $\epsilon$, one recovers Eq.~(\ref{eq:density_multidim}).

\subsection{Scaling of $\aal$ and $N$ close to the critical point}
During our discussion, we have always supposed the limit $N\to\infty$ to be attained before any other one, notably the one $\aal\to 1$ corresponding to the critical regime of a linear system. In this last part of the appendix, we want to show that it is possible to relax this assumption, and provided that $\aal$ thends to one slower than
\begin{equation}
  \label{eq:scaling_alpha_n}
  \aal = 1 - \frac{K}{N} \;,
\end{equation}
we obtain a finite limit for the auto-correlations $\avcd{c(\tau)}$. In fact, for $\lambda_\veca $ real, we can always write Eq.~(\ref{eq:acorr_factor})
\begin{eqnarray}
  \label{eq:bound_corr}
  \avcd{c(\tau)} &\leq& \avcd{c(0)} \leq \frac{ \kappa}{L^D} \sum_{\veca \in \{0,\dots,L-1\}^D}  \frac{1 }{(1-\lambda_a)} \\
&=& \frac{ \kappa}{L^D} \sum_{\veca \in \{0,\dots,L-1\}^D}  \frac{1 }{1-\aal D^{-1}\sum_{d=1}^D\cos(2\pi a_d/L)}
\end{eqnarray}
where $\kappa$ is a constant equal to $\beta \bar \Sigma$. The terms in which the cosine is negative can be bound by $1$, and thus give a finite contribution. By symmetry of the cosine function, one can  additionally restrict the remaining terms to the subset  $\veca \in \{0,\dots,L/4\}^D$, where $\cos(x) \leq 1-Cx^2$, and $C=1/3$. It results that
\begin{equation}
   \avcd{c}(\tau) \leq \frac{ \kappa^\prime}{L^D} \sum_{\veca \in \{0,\dots,L/4\}^D}  \frac{1}{1-\aal D^{-1}\sum_{d=1}^D(1-4\pi^2 a^2_d/L^2)} \;.
\end{equation}
The potential divergence at small $\veca$ comes only from the $\veca=\vecz$ term, and can be reabsorbed by a scaling of $\aal$ slower than $\aal = 1-K/L^D$, in whose case one can safely approximate the above sum by an integral. Such integral can be bound by
\begin{equation}
   \avcd{c}(\tau) \leq \frac{\kappa^{\prime\prime}}{L^D} \int_{1}^{L/4} dr \frac{r^{D-1}}{(1-\aal) +4\pi^2\aal \frac{r^2}{L^2}} \;,
\end{equation}
where $\kappa^{\prime\prime}$ is a constant taking into account an integration over the angular components arising from a transformation in polar coordiantes and the small $r$ contribution to the sum. This integral converges for $D>2$ even for $\aal=1$. Summarizing, provided that the divergence at small distances is tamed by an appropriate scaling of $\aal$, the large length behavior is dictated only by dimensionality, regardless of the scaling of $\aal$.

\section{Observables of a random linear system}
\label{app:obs}
In Sec.~\ref{sec:goe} we have been interested interested in calculating the momenta of several observables (the correlations $\hat c_{ij}(\omega)$, the mean intensities $\Lambda_i$ or the exogenous intensities $\mu_i$) according to the expectation $\avs{\dots} $ associated with the GOE statistical ensemble. In this appendix, we will describe the systematic method which has been developed in order to calculate all these quantities.
As a preliminary note, we remark that in all such ensembles, the momenta of $\Lambda_i$ and $c_{ij}(\tau)$, are by construction invariant under permutation of the indices. That is why, without loss of generality, complete information about the system can be encoded in the scalar expressions appearing in Sec.~\ref{sec:obs_goe}.
For brevity, we list in this appendix just the results that we get for the  intensities $\avs{\Lambda_i}$ and the average correlations $\avs{\hat c_{ij}(\omega)}$ in the VAR case with random drive. In fact, the procedure that we illustrate can be generalized straightforwardly to the other scenarios, leading to the expressions reported in  Sec.~\ref{sec:obs_goe}.

First, in the VAR case we remark that the decomposition~(\ref{eq:decomp_goe}) leads to an the expression for $\hat \Psi(\omega) =(\id - \hat \Phi(\omega))^{-1}$ of the type
\begin{equation}
\hat \Psi(\omega) = \hat\phi(\omega)^{-1} O \left[ (\id\hat\phi(\omega)^{-1} - \lambda )^{-1} \sum_{n=0}^\infty \left( \frac{\aal}N \right)^n \left( (O^T \one O)(\id \hat\phi(\omega)^{-1} - \lambda )^{-1}\right)^n \right] O^T \; .
\end{equation}
This allows to write explicitly the expression for the vector $\Lambda$ and the correlation matrix $\hat c(\omega)$, which become
\begin{eqnarray}
\Lambda &=& \hat \Psi(0) \mu = \sum_n \left( \frac{\aal}{N} \right)^n O \hat K(0) [M \hat K(0)]^n O^T \\
\hat c(\omega) &=&  || \hat \phi(\omega)^{-1}||^2 \sum_{n,m} \left( \frac{\aal}{N} \right)^{n+m} O \hat K^\star(\omega) [M \hat K^\star(\omega)]^n O^T \asi O  [\hat K(\omega) M]^m \hat K(\omega) O^T \, , \nonumber \\
&&
\end{eqnarray}
where $M$ is the symmetric matrix $M=O^T \one O$ and where $\hat K(\omega) = (\id\hat\phi(\omega)^{-1} - \lambda )^{-1}$. The relevant scalars for the averages of these quantities become
\begin{eqnarray*}
 \avs{ \avcd{ \Lambda}} &=& \frac{\amu}{N} \sum_n \left( \frac \aal N \right)^n  \avs{\tr \left\{ (M \hat K_0)^{n+1} \right\}} \\
\avs{\avcd{ \hat c(\omega)}} &=& \frac \asi N || \hat \phi(\omega)||^{-2} \sum_{n,m}  \left( \frac{\aal}{N} \right)^{n+m} \avs{\tr \left\{
\hat K^\star(\omega) (M \hat K^\star(\omega))^n  (\hat K(\omega) M)^m \hat K(\omega)
\right\} }\\
\avs{\avcm{ \hat c(\omega)}} &=& \frac \asi {N^2} || \hat \phi(\omega)||^{-2} \sum_{n,m}  \left( \frac{\aal}{N} \right)^{n+m} \avs{\tr \left\{
\hat K^\star(\omega) (M \hat K^\star(\omega))^n  (\hat K(\omega) M)^m \hat K(\omega) M
\right\} } \; .
\end{eqnarray*}
These expressions are more conveniently computed after introducing the generating function
\begin{equation}
F^{(p)}(\vec x,\{z_a^{(k)}\}_{a=1}^N) = \sum_{n^{(1)}} \dots \sum_{n^{(p)}} \prod_{k=1}^p \left\{ \frac{x^{(k)}}{N} \sum_a z_a^{(k)} \left( \sum_i O_{ia}\right)^2 \right\}^{n^{(k)}} \; ,
\end{equation}
which satisfies the identities:
\begin{eqnarray}
F^{(p)}(0,\{z_a^{(k)}\}_{a=1}^N)  &=& 1 \\
\frac{1}{n!}\frac{ \partial^{n} }{\partial (x^{(k)})^n} F^{(p)}(\vec x,\{z_a^{(k)}\}_{a=1}^N) \Bigg|_{x^{(k)}=0} &=&  \left\{ \frac{1}{N} \sum_a z_a^{(k)}\left( \sum_i O_{ia}\right)^2 \right\}^n\\
&\times &  F^{(p-1)}(x^{(1)},\dots,x^{(k-1)},x^{(k+1)},\dots,x^{(p)}, \nonumber \\
& & \{z_a^{(1)}\},\dots, \{z_a^{(k-1)}\}, \{z_a^{(k)}\},\dots, \{z_a^{(p)}\})   \nonumber \; .
\end{eqnarray}
By using these function, it is not hard to show that all the trace expressions can be computed as derivatives with respect to the $x^{(k)}$ variables. Indeed, in order to recover the expression listed in Sec~\ref{sec:obs_goe} it's still necessary to average the generating function $F^{(p)}(\vec x,\{z_a^{(k)}\}_{a=1}^N)$ over the distribution of the symmetric matrices $O$ over the Haar measure. In order to do that, we need to use a result proved in the next appendix, where it is shown that
\begin{equation}
\avs{ \prod_a \left(\sum_i O_{ia} \right)^{2n_a}} = N^n \frac{\Gamma(N/2)}{4^n \Gamma(N/2+n)} \prod_a \frac{(2n_a)!}{n_a!} \; ,
\end{equation}
where $n=\sum_a n_a$.
Then, we can expand $\avs{F^{(p)}(\vec x,\{z_a^{(k)}\}_{a=1}^N)}$ as
\begin{eqnarray*}
\avs{F^{(p)}(\vec x,\{z_a^{(k)}\}_{a=1}^N)} &=& \avs{\sum_{ n^{(1)}} \dots \sum_{ n^{(p)}} \prod_{k=1}^p \left\{ \frac{x^{(k)}}{N} \sum_a z_a^{(k)} \left( \sum_i O_{ia}\right)^2 \right\}^{n^{(k)}} }\\
&=& \overline{\mathbb{E}} \left[ \sum_{\vec n^{(1)}} \dots \sum_{\vec n^{(p)}} \prod_{k=1}^p \prod_{a=1}^N \left[ \left( \frac{x^{(k)} z_a^{(k)}}{4} \right)^{n_a^{(k)}} \frac{1}{n_a^{(k)}!} \right]  \right.\\
&\times& \left. \frac{\Gamma(N/2)}{\Gamma(N/2+\sum_k n^{(k)})} \prod_k \Gamma(n^{(k)}+1) \frac{(2\sum_k n^{(k)})!}{(\sum_k n^{(k)})!} \right] \; ,
\end{eqnarray*}
where the $\vec n^{(k)} = (n^{(k)}_1,\dots,n^{(k)}_N )$ variables have been introduced by expanding the multinomials appearing in the first line. Then we can expand the products of Gamma functions as products of Beta functions by using
\begin{eqnarray*}
&& \frac{\Gamma(N/2)}{\Gamma(N/2+\sum_k n^{(k)})} \prod_k \Gamma(n^{(k)}+1) \\
&=& \frac{\Gamma(N/2)}{\Gamma(N/2 - p)} B(n^{(1)}+1,n^{(2)}+1) B(n^{(1)} +n^{(2)}+2,n^{(3)}+1) \dots \, B\left(\sum_k n^{(k)} + p,N/2-p\right) \\
&=& \int_0^1 dt_1 \dots dt_p  \;, t_1^{n^{(1)}} (1-t_1)^{n^{(2)}} t_2^{n^{(1)} +n^{(2)} +1} (1-t_2)^{n^{(3)}} \dots \, t_p^{\sum_k n^{(k)} + p-1} (1-t_p)^{N/2-p-1}
\end{eqnarray*}
in order to write
\begin{eqnarray*}
\avs{F^{(p)}(\vec x,\{z_a^{(k)}\}_{a=1}^N)}
&=& \frac{\Gamma(N/2)}{\Gamma(N/2-p)} \int_0^1 dt_1 \dots dt_p \prod_{k=1}^p t_k^{k-1} (1-t_p)^{N/2-p-1} \nonumber \\
&\times & \sum_{\vec n^{(1)}} \dots \sum_{\vec n^{(p)}}  \prod_{k=1}^p \prod_{a=1}^N \left[ \left( \frac{x^{(k)} h^{(k)}(\vec t) z_a^{(k)}}{4} \right)^{n_a^{(k)}} \frac{1}{n_a^{(k)}!} \right]
 \frac{(2\sum_k n^{(k)})!}{(\sum_k n^{(k)})!} \; ,
\end{eqnarray*}
where the functions $h^{(k)}(\vec t)$ are defined as in Sec.~\ref{sec:obs_goe}.
Finally, one needs to employ recursively the identity
\begin{equation}
\sum_{n=0}^{\infty} t^n \frac{(2(n+m))!}{n! (n+m)!} \frac{(2m)!}{m!} (1-4t)^{-1/2 -m}
\end{equation}
in order to sum the series appearing in the above expression for $\avs{F^{(p)}(\vec x,\{z_a^{(k)}\}_{a=1}^N)}$ and obtain
\begin{eqnarray}
\avs{F^{(p)}(\vec x,\{z_a^{(k)}\}_{a=1}^N)} &=&  \frac{\Gamma(N/2)}{\Gamma(N/2-p)} \int_0^1 dt_1 \dots dt_p \prod_{k=1}^p t_k^{k-1} (1-t_p)^{N/2-p-1} \nonumber \\
&\times & \prod_a \left(1 - \sum_{k=1}^p h^{(k)}(\vec t) x^{(k)} z_a^{(k)} \right)^{-1/2} \nonumber \\
&=& \gf{p} (\vec x,\{z_a^{(k)}\}_{a=1}^N)   \; .
\end{eqnarray}
You can notice that all the observables listed in Sec.~\ref{sec:obs_goe} are conveniently expressed in terms of the derivatives of $ \gf{p} (\vec x,\{z_a^{(k)}\}_{a=1}^N)  $.
We finally remark that:
\begin{itemize}
\item
Finding the derivatives $ \gf{p} (\vec x,\{z_a^{(k)}\}_{a=1}^N)$ calculated in $\vec x=0$ is straightforward, as the integrals in the $\vec t$ variables factorize. This property can be used to find systematically the terms of a small $\aal$ expansion for the expression of the observables, and in particular has been used in order to write down the $\aal=0$ expressions listed in Sec.~\ref{sec:obs_goe}.
\item
The leading term of the large $N$ expansion of $ \gf{p} (\vec x,\{z_a^{(k)}\}_{a=1}^N)$ is particularly simple, and results:
\begin{equation}
\label{eq:gf_inf}
\gf{p} (\vec x,\{z_a^{(k)}\}_{a=1}^N) = \prod_k \left(1-\frac \aal N \sum_a z_a^{(k)} \right)^{-1} \; ,
\end{equation}
indicating that, provided that the $ z_a^{(k)}$ variables are of order 1, the generating function is also of order 1.
\item
In the degenerate case in which all the $z_a$ are equal to $z$, then the generating function $\gf{p} (\vec x,\{z^{(k)}\}_{a=1}^N)$ is \emph{exactly} (i.e., to all orders in $N$) equal to
\begin{equation}
  \label{eq:gf_homog}
  \gf{p} (\vec x,\{z^{(k)}\}_{a=1}^N) = \prod_k \left(1-\aal z^{(k)} \right)^{-1}
\end{equation}

\end{itemize}

\section{Momenta over the Haar sphere}
\label{app:haar}
In this appendix want to compute the momenta of random symmetric matrices $O$ distributed according to the Haar measure associated to the group $O(N)$. In order to do that, we can exploit the Weingarten formalism~\cite{Collins:2006aa,Collins:2009aa}, which is used to compute generic expressions of the type
\begin{equation}
\av{O_{i_1 a_1} \dots O_{i_{2n}a_{2n}} } = \sum_{p,q\in {\mathcal P}_2(2n)} \delta_{i_1 i_{p(1)}} \dots \delta_{i_{2n} i_{p(2n)}} \delta_{a_1 a_{q(1)}} \dots \delta_{a_{2n} a_{q(2n)}} W^{O}(p,q)
\end{equation}
where ${\mathcal P}_2(2n)$ is the set of all possible pairings of $\{1,\dots,2n\}$ and $W^O(p,q)$ is the element $(p,q)$ of the orthogonal Weingarten matrix ~\cite{Collins:2006aa,Collins:2009aa}.
In particular, we want to exploit this formula in order to compute the expectation
\begin{eqnarray}
&&\av{ \prod_a \left(\sum_i O_{ia} \right)^{2n_a} } \\
&=& \sum_{i_1, j_1} \dots \sum_{i_n j_n}  \av{
\underbrace{O_{i_1 1} O_{j_1 1} \dots  O_{i_{n_1} 1} O_{j_{n_1} 1} }_{n_1 \textrm{ terms}}
\dots
\underbrace{O_{i_{n-n_N+1} N} O_{j_{n-n_N+1} N} \dots O_{i_n N} O_{j_n N}  }_{n_N \textrm{ terms}}
} \nonumber \\
&=& N^n  \sum_{p,q\in {\mathcal P}_2(2n)} \delta_{a_1 a_{q(1)}} \dots \delta_{a_{2n} a_{q(2n)}} W^{O}(p,q)  \label{eq:momenta} \; .
\end{eqnarray}
Its value can be obtained by solving the simpler problem of computing momenta of the components of a random vector $\vec O_i=(O_{i1},\dots,O_{iN})$ distributed uniformly on the $N$-dimensional unit sphere. The expression for its momenta is in fact proportional to Eq.~(\ref{eq:momenta}):
\begin{equation}
\av{ \prod_a (O_{ia})^{2n_a} } =  \sum_{p,q\in {\mathcal P}_2(2n)} \delta_{a_1 a_{q(1)}} \dots \delta_{a_{2n} a_{q(2n)}} W^{O}(p,q) \label{eq:momenta_simple}\; .
\end{equation}
Hence, once the value of Eq.~(\ref{eq:momenta_simple}) has been computed, the value of Eq.~(\ref{eq:momenta}) can be recovered after multiplying by $N^n$.

The value of Eq.~(\ref{eq:momenta_simple}) is easily obtained by constructing the generating function
\begin{equation}
Z[\vec J] = \int d\vec u \, e^{\vec J \cdot \vec u} \; \delta(|\vec u|^2 -1) \; ,
\end{equation}
which can be evaluated exponentiating the delta function:
\begin{eqnarray}
Z[\vec J] &\propto& \int \frac{dz}{2\pi} d\vec u \; e^{\sum_{a=1}^N J_a  u_a - i z (\sum_i u_a^2 -1 )} \\
&=&  \int_{-\infty}^{+\infty} \frac{dz}{2\pi} e^{iz} \prod_a \left(  \int_{-\infty}^{+\infty}  du_a e^{J_a u_a - iz u_a^2} \right) \\
&=&   \int_{-\infty}^{+\infty} \frac{dz}{2\pi} e^{iz} \prod_a e^{\frac{J_a^2}{4iz}} \left( \frac{\pi}{iz} \right)^{1/2} \\
&=&  \int_{-\infty}^{+\infty} \frac{dz}{2\pi} e^{iz}  \sum_{n=0}^\infty \frac{(\sum_a J_a^2/4)^n}{n!} \frac{ \pi^{N/2}}{(iz)^{n+N/2}} \\
&=&  \pi^{N/2}\sum_{n=0}^\infty \frac{(\sum_a J_a^2/4)^n}{n!}  \int_{-\infty}^{+\infty} \frac{dz}{2\pi} e^{iz} \frac{1}{(iz)^{n+N/2}} \\
&=&  \frac{\pi^{N/2}}{2} \sum_{n=0}^\infty \frac{(\sum_i J_a^2/4)^n}{\Gamma(n+1) \Gamma(n+N/2)} \; ,
\end{eqnarray}
One gets finally
\begin{equation}
Z[\vec J]  \propto   2 \pi^{N/2} \sum_{n=0}^\infty \frac{x^n}{\Gamma(n+1) \Gamma(n+N/2)} \; ,
\end{equation}
with $x = \sum_a J_a^2/4$.

By using this result, it is easy to calculate the average components of a vector uniformly distributed on the unit sphere:
\begin{eqnarray}
\frac{1}{Z(0)} \frac{\partial^{2n}  Z(\vec J) }{\partial J_{1}^{2 n_1} \dots \partial J_{N}^{2n_N}}\Bigg|_{\vec J =0} &=& \av{(O_{i 1})^{2n_1} \dots (O_{i N})^{2n_N}} \\ \nonumber
&=& \frac{\Gamma(N/2)}{4^n \Gamma(N/2 + n )} \prod_{p=1}^m \frac{(2n_p)!}{n_p!} \; ,
\end{eqnarray}
where $n=\sum_{a=1}^N n_a$.

\bibliography{random_linear_processes}{}
\bibliographystyle{unsrt}

\end{document}